\documentclass[a4paper,onecolumn,11pt,accepted=2026-01-07]{quantumarticle}
\pdfoutput=1
\usepackage[utf8]{inputenc}
\usepackage[english]{babel}
\usepackage[T1]{fontenc}
\usepackage{amsmath}
\usepackage{amssymb}
\usepackage{amsfonts}
\usepackage{hyperref}
\usepackage[numbers]{natbib}

\usepackage{tikz}
\usepackage{lipsum}

\usepackage[utf8]{inputenc}
\usepackage{graphicx}
\usepackage{color}
\usepackage{physics}
\usepackage{braket}
\usepackage{comment}
\usepackage{mathrsfs}
\usepackage{enumerate}

\usepackage[version=4]{mhchem}
\usepackage[breaklinks=true]{hyperref}
\usepackage{lipsum}
\usepackage{algorithmic}
\usepackage{algorithm}

\newcommand{\mr}[1]{\mathrm{#1}}

\newcommand{\mcl}[1]{\mathcal{#1}}

\newcommand{\bbR}{\mathbb{R}}

\newcommand{\bbN}{\mathbb{N}}

\newcommand{\ad}{\mathrm{ad}}

\newcommand{\poly}[1]{\mathrm{poly} \left( #1 \right)}

\usepackage{amsthm}
\theoremstyle{definition}
\newtheorem{theorem}{Theorem}[]

\newtheorem{lemma}[theorem]{Lemma}

\newtheorem*{theorem*}{Theorem}
\newtheorem*{proposition*}{Proposition}

\begin{document}

\title{On the commutator scaling in Hamiltonian simulation with multi-product formulas}

\author{Kaoru Mizuta}
\email{mizuta@qi.t.u-tokyo.ac.jp}
\affiliation{Department of Applied Physics, Graduate School of Engineering, The University of Tokyo, Hongo 7-3-1, Bunkyo, Tokyo 113-8656, Japan}
\affiliation{Photon Science Center, Graduate School of Engineering, The University of Tokyo, Hongo 7-3-1, Bunkyo, Tokyo 113-8656, Japan}
\affiliation{RIKEN Center for Quantum Computing (RQC), Hirosawa 2-1, Wako, Saitama 351-0198, Japan}
\orcid{0009-0004-6156-1984}
\maketitle

\begin{abstract}
  A multi-product formula (MPF) is a promising approach for Hamiltonian simulation efficient both in the system size $N$ and the inverse allowable error $1/\varepsilon$ by combining Trotterization and linear combination of unitaries (LCU).
The efficiency in $1/\varepsilon$ comes from the Richardson extrapolation with the well-conditioned coefficients, achieving poly-logarithmic cost in $1/\varepsilon$ like LCU [G. H. Low, V. Kliuchnikov, N. Wiebe, arXiv:1907.11679 (2019)].
The efficiency in $N$ is expected to come from the commutator scaling in Trotterization, and this appears to be confirmed by the error bound of MPF expressed by nested commutators [J. Aftab, D. An, K. Trivisa, arXiv:2403.08922 (2024)].
However, we point out that the efficiency of MPF in the system size $N$ is not exactly resolved yet in that the present error bound expressed by nested commutators is incompatible with the size-efficient complexity reflecting the commutator scaling.
The problem is that $q$-fold nested commutators with arbitrarily large $q$ are involved in their requirement and error bound.
The benefit of commutator scaling by locality is absent, and the cost efficient in $N$ becomes prohibited in general.
In this paper, we show an alternative commutator-scaling error of MPF and derive its size-efficient cost properly inheriting the advantage in Trotterization.
The requirement and the error bound in our analysis, derived by techniques from the Floquet-Magnus expansion, have a certain truncation order in the nested commutators and can fully exploit the locality.
We prove that Hamiltonian simulation by MPF certainly achieves the cost whose size-dependency is as large as Trotterization while keeping the $\mr{polylog}(1/\varepsilon)$-scaling like LCU. 
Our results will provide improved or accurate error and cost also for various algorithms using interpolation or extrapolation of Trotterization.
\end{abstract}

\section{Introduction}

Simulating the dynamics of quantum many-body systems, called Hamiltonian simulation, is one of the significant applications of quantum computers with promising quantum speedup.
The ultimate goal in this field is to establish an optimal quantum algorithm with smaller computational cost in the system size $N$, the simulation time $t$, and the inverse allowable error $1/\varepsilon$.
Trotterization, or called product formulas (PFs), is one of the most famous approaches, featuring simple circuit structures and requiring no ancilla qubits \cite{Suzuki1991-za,Lloyd1996-ko}.
In addition, it has favorable gate complexity in the system size $N$ due to the commutator-scaling error \cite{Somma2015-fs-commutator,childs-prl2019-pf,childs2021-trotter}.
Post-Trotter approaches, such as linear combination of unitaries (LCU) and quantum singular-value transform (QSVT), are alternative quantum algorithms that operate with oracles calling Hamiltonians \cite{Berry-prl2015-LCU,Low2017-qsp,Low2019-qubitization,Gilyen2019-qsvt}.
They have exponentially better cost in the inverse allowable error $1/\varepsilon$ than Trotterization owing to the usage of accurate polynomial approximation of time-evolution operators.
On the other hand, their scaling in the system size $N$ is less favorable due to the oracle overhead.

The central question in Hamiltonian simulation has been how we can construct efficient quantum algorithms both in the system size and the accuracy.
A multi-product formula (MPF) \cite{Childs_Wiebe_2012_mpf,low2019-mpf,Faehrmann_2022-mpf,Vazquez2022-mpf,Carrera_Vazquez_2023-mpf,zhuk2024-mpf,aftab2024-mpf} is a promising approach, which is a linear combination of Trotterization with different time steps.
Coefficients and time steps in the linear combination based on the Richardson extrapolation allows MPF to cancel some of dominant Trotter errors, giving higher order approximation of the time-evolution operator.
In particular, the well-conditioned MPF, in which the low-weight coefficients and time steps are carefully chosen, can be efficiently implemented by the LCU approach \cite{low2019-mpf}.
It has been shown that the well-conditioned MPF can achieve $\mr{polylog}(1/\varepsilon)$ computational cost like the LCU approach \cite{low2019-mpf}.
The efficiency of MPF in the system size has also been vigorously explored via the analysis of its error bound \cite{zhuk2024-mpf,aftab2024-mpf}.
MPF is expected to inherit the commutator scaling in Trotterization and thereby host substantial reduction in the error and complexity owing to the locality of Hamiltonians.
Indeed, Ref. \cite{aftab2024-mpf} has recently derived the error bound of generic MPF explicitly expressed by nested commutators.
Based on this error bound, it concluded that the size-dependence of the cost of MPF is comparable to that of Trotterization and better than the post-Trotter methods like LCU. 
Recent studies seem to settle the issues surrounding the MPF.
Namely, Hamiltonian simulation based on MPF is efficient both in the system size and the allowable error with reflecting the advantages of Trotterization and LCU.

However, focusing on the efficiency of MPF with respect to the system size, we point out that the existing error bound of MPF expressed by nested commutators \cite{aftab2024-mpf} does not imply the size-efficient cost reflecting the commutator scaling.
The problem is that $q$-fold nested commutators with arbitrarily large $q$ are contained in the requirement and the error bound.
This comes from the necessity of the convergence of the Baker-Campbell-Hausdorff (BCH) formula in the derivation (See Section \ref{Subsec:Problem} in detail).
When the order $q$ can become arbitrarily large, the substantial reduction of the nested commutators due to the locality of Hamiltonians is no longer present.
As a result, the computational cost generally obeys the $1$-norm scaling, whose size-dependency is larger than the one of the commutator scaling.
Namely, it remains unresolved yet whether MPF can achieve the size-efficient cost reflecting the commutator scaling.

In this paper, we provide a complete answer to it; we derive an alternative error bound expressed by nested commutators for generic MPF and prove its proper size-efficient cost reflecting the commutator scaling.
Our error analysis includes nested commutators up to a certain truncation order in its requirement and error bound.
The problem around the BCH formula is resolved by a technique developed for the Floquet-Magnus expansion \cite{Abanin2015-zg,Kuwahara2016-yn,Mori2016-tp,Abanin2017-li,Abanin2017-zs}.
The truncation order makes the locality of Hamiltonians work well, resulting in a much looser requirement and a better error bound compared to the previous analysis \cite{aftab2024-mpf}.
Based on this error bound, we establish the proper cost whose size-dependency is comparable to that of Trotterization while keeping its polylogarithmic dependency in $1/\varepsilon$.
Our results conclude that Hamiltonian simulation by MPF is efficient with respect to both the system size and the allowable error.
Namely, the MPF certainly inherits the advantages respectively from Trotterization and LCU.
There are various quantum algorithms sharing the mathematical background with MPF from the viewpoint of interpolation or extrapolation for Trotterization \cite{Endo-PRA2019-mitigation,Rendon2023-interpolate,Rendon2024-interpolate,Watson2024-mpf,watson2024-mpf-random,chakraborty2025-mpf,Guo2025-interpolate}.
We expect that our approach will give a precise analysis of their errors and cost in the size-dependency, reflecting the commutator scaling.

\section{Preliminary}\label{Sec:Preliminary}

In Section \ref{Subsec:Setup}, we specify the setup and review the multi-product formula (MPF).
In Section \ref{Subsec:Problem}, we discuss the error bound on MPF, especially focusing on the problem of the commutator-scaling error in Ref. \cite{aftab2024-mpf}.
In Section \ref{Subsec:Summary}, we summarize our results on the modified error bound and cost of MPF and highlight the differences from Ref. \cite{aftab2024-mpf}.

\subsection{Setup and multi-product formula (MPF)}\label{Subsec:Setup}

Throughout this paper, we consider a system on an $N$-qubit lattice $\Lambda = \{1,\cdots,N\}$.
We assume the $k$-locality of the Hamiltonian $H$ by
\begin{equation}\label{Eq_Pre:Hamiltonian}
    H = \sum_{X\subset \Lambda: |X| \leq k} h_X,
\end{equation}
where each $h_X$ denotes a local term acting on a domain $X$ and satisfies $[h_X,h_{X'}]=0$ if $X\cap X' = \phi$.
We assume $k \in \order{1}$ throughout the paper.
We also define the $g$-extensiveness by
\begin{equation}\label{Eq_Pre:extensiveness}
    \sum_{X: X \ni j} \norm{h_X} \leq g, \quad ^\forall j \in \Lambda,
\end{equation}
where $\norm{\cdot}$ denotes the operator norm.
The extensiveness $g$ means the maximum energy per site.
It depends on the range of interactions as
\begin{equation}
    g \in \begin{cases}
        \order{N^0} & (\text{finite-range interactions}), \\
        \order{N^0} & (\text{long-range interactions, $\nu>d$}), \\
        \order{\log N} & (\text{long-range interactions, $\nu=d$}), \\
        \order{N^{1-\nu/d}} & (\text{long-range interactions, $\nu<d$}),
    \end{cases}
\end{equation}
where $d$ denotes the spatial dimension of the lattice $\Lambda$.
The quantity $\nu$ means the power of the long-range interactions, where the distance-$r$ interactions are assumed to decay as $\order{r^{-\nu}}$.

The goal of Hamiltonian simulation is to find an operator that accurately approximates the time evolution operator $e^{-iHt}$ and can be efficiently implemented on quantum circuits.
Trotterization, or product formulas (PFs), is one of the most common approaches.
Suppose that we decompose the Hamiltonian $H$ by
\begin{equation}\label{Eq_Pre:H_partition}
    H = \sum_{\gamma=1}^\Gamma H_\gamma, \quad H_\gamma = \sum_X h_X^\gamma.
\end{equation}
We use the notation $h_X$ in Eq. (\ref{Eq_Pre:Hamiltonian}) such that the collection $\{ h_X^\gamma \}_{X,\gamma}$ comprises the set of terms $\{h_X\}_X$, where each $h_X^\gamma$ corresponds to one element of $\{h_X\}$.
With this notation, the extensiveness by Eq. (\ref{Eq_Pre:extensiveness}) reads
\begin{equation}
    \sum_{\gamma=1}^\Gamma \sum_{X: X \ni j} \norm{h_X^\gamma} \leq g, \quad ^\forall j \in \Lambda.
\end{equation}
The $p$-th order formula $T_p(\tau)$ for small time $\tau$ is defined by
\begin{equation}\label{Eq_Pre:Trotterization}
    T_p(\tau) = \prod_{v=1,\cdots,c_p\Gamma}^\leftarrow e^{-iH_{\gamma_v} \alpha_v \tau}, \quad \prod_{v=1,\cdots,V}^\leftarrow U_v \equiv U_V \cdots U_2 U_1,
\end{equation}
where the coefficients $\{\alpha_v\}$ are determined so that it satisfies the order condition, $T_p(\tau)=e^{-iH \tau}+\order{\tau^{p+1}}$ under $\tau \to 0$.
The number $c_p$ denotes the repetition, which usually increases exponentially in $p$.
Due to this, we focus on the case with the order $p \in \order{1}$ [and hence $c_p$ is also $\order{1}$].
The number $\Gamma$ depends on Hamiltonians, and it can be $\order{1}$ for finite-ranged interacting systems while it generally belongs to $\poly{N}$ for long-ranged interacting systems (See Ref. \cite{mizuta2025_low_energy} for example).
For example, the first- and second-order formulas are given by $T_1(\tau) = \prod_{\gamma=1,\cdots,\Gamma}^\leftarrow e^{-iH_\gamma \tau}$ and $T_2(\tau)=T_1(-\tau/2)^\dagger T_1(\tau/2)$.
Generic Trotter formulas exhibit a so-called commutator-scaling error \cite{childs2021-trotter}, expressed as
\begin{equation}\label{Eq_Pre:error_trotter}
    \norm{e^{-iH\tau}-T_p(\tau)} \in \order{\alpha_{\mr{com},p+1}\tau^{p+1}},
\end{equation}
\begin{equation}\label{Eq_Pre:commutator_bound}
    \alpha_{\mr{com},q} \equiv \sum_{\gamma_1,\cdots,\gamma_q=1}^\Gamma \norm{[H_{\gamma_q},\cdots,[H_{\gamma_2},H_{\gamma_1}]]} \leq (q-1)!(2kg)^{q-1} Ng. 
\end{equation}
This is the origin of the fact that Trotterization can achieve better cost in the system size $N$ than other approaches like LCU and QSVT \cite{Berry-prl2015-LCU,Low2019-qubitization,Gilyen2019-qsvt}.

In contrast to the better cost in the system size $N$ owing to the commutator scaling, the bottleneck of Trotterization is its enormous cost in accuracy, which increases polynomially in the inverse allowable error $1/\varepsilon$.
One of the solutions to this is to combine it with the LCU approach that can achieve exponentially better cost in $1/\varepsilon$, which leads to the multi-product formula (MPF).
We define MPF with the orders $p,m,J \in \bbN$ by a linear combination of the $p$-th order Trotterization,
\begin{equation}\label{Eq_Pre:MPF}
    M_{pmJ}(\tau) = \sum_{j=1}^J c_j \left[ T_p(\tau/k_j) \right]^{k_j},
\end{equation}
where the numbers $c_j \in \bbR$ and $k_j \in \bbN$ are organized so that it can satisfy the order condition, 
\begin{equation}\label{Eq_Pre:MPF_order}
    M_{pmJ}(\tau) = e^{-iH\tau} + \order{\tau^{m+1}}.
\end{equation}
One of the ways to determine $\{c_j\}$ and $\{k_j\}$ relies on the Richardson extrapolation \cite{Chin2010-mpf,Endo-PRA2019-mitigation,Carrera_Vazquez_2023-mpf}.
For an even order $p$ satisfying the symmetric condition $T_p(-\tau)^\dagger=T_p(\tau)$, the solution to
\begin{eqnarray}
   \left( \begin{array}{cccc}
        1 & 1 & \cdots & 1 \\
        (k_1)^{-2} & (k_2)^{-2} & \cdots & (k_J)^{-2} \\
        \vdots & \vdots & & \vdots \\
        (k_1)^{-2J+2} & (k_2)^{-2J+2} & \cdots & (k_m)^{-2J+2}
    \end{array}\right)
    \left( \begin{array}{c}
         c_1 \\
         c_2 \\
         \vdots \\
         c_J
    \end{array}\right)
    =
    \left( \begin{array}{c}
         1 \\
         0 \\
         \vdots \\
         0
    \end{array}\right)
\end{eqnarray}
with an arbitrary set of different natural numbers $\{k_j\}$ gives a MPF satisfying $M_{pmJ}(\tau)=e^{-iH\tau}+\order{\tau^{2J+1}}$ (i.e., $m=2J$).
While there are still several candidates due to the choice of $\{k_j\}$, the solution should satisfy 
\begin{equation}\label{Eq_Pre:well_condition}
    \norm{k}_1 \equiv \sum_{j=1}^J |k_j| \in \poly{J}, \quad \norm{c}_1 \equiv \sum_{j=1}^J |c_j| \in \poly{J}
\end{equation}
to efficiently implement Eq. (\ref{Eq_Pre:MPF}) by the LCU approach \cite{Berry-prl2015-LCU}.
Such a solution is called the well-conditioned solution, and Ref. \cite{low2019-mpf} has found the one satisfying $\norm{k}_1 \in \order{J^2 \log J}$ and $\norm{c}_1 \in \order{\log J}$, which is valid for generic even $p$.

\subsection{Error and complexity of MPF and the problem of the existing analysis}\label{Subsec:Problem}

The error bound of MPF determines its complexity, i.e., the Trotter number as well as the Trotterization.
For simulating the quantum dynamics over the large time $t$, we split it into $r$ parts and set the small time $\tau = t/r$.
The Trotter number $r$ is determined so that the error per step can be bounded by
\begin{equation}\label{Eq_Pre:r_condition}
    \norm{e^{-iHt}-\left[ M_{pmJ}(t/r)\right]^r} \leq \varepsilon.
\end{equation}
In the LCU for implementing MPF, we call a controlled Trotterization operator,
\begin{equation}
    C[T_p(t/(rk_j))] = \ket{0}\bra{0} \otimes I + \ket{1}\bra{1} \otimes T_p(t/(rk_j)).
\end{equation}
The number of queries to $C[T_p(\tau)]$ with some time $\tau$ amounts to $\order{\norm{c}_1 \norm{k}_1 r}$, where $\norm{k}_1$ means the repetition number in each step $t/r$ and the factor $\norm{c}_1$ comes from the quantum amplitude amplification (QAA) \cite{Brassard_2002_qaa} to execute the LCU approach with certainty.
Using the well-conditioned solution, Eq. (\ref{Eq_Pre:well_condition}), MPF has been shown to have polylogarithmic query complexity in the inverse allowable error $1/\varepsilon$ based on the order condition Eq. (\ref{Eq_Pre:MPF_order}) \cite{low2019-mpf}.

By contrast, the efficiency in the system size $N$ is rather complicated since we have to determine the size-dependency of the error term $\order{\tau^{m+1}}$ in Eq. (\ref{Eq_Pre:MPF_order}).
Ref. \cite{low2019-mpf} showed that the error bound can be expressed by the nested commutators among $\{ H_\gamma \}$, but its explicit form was not determined.
Ref. \cite{zhuk2024-mpf} derived the commutator-scaling error of MPF, but it was restricted to the case $J=p+1$ and not applicable to the well-conditioned MPF.
Ref. \cite{aftab2024-mpf} gave the following error bound applicable to generic well-conditioned MPFs.

\begin{theorem}\label{Thm:original}
\textbf{(Theorem in Ref. \cite{aftab2024-mpf})}

Suppose that the time $\tau$ is small enough to satisfy
\begin{equation}\label{Eq_Pre:requirement_original}
    ^\exists q_0>p, \quad \text{s.t.} \quad \alpha_{\mr{com},q} \tau^q < 1, \quad ^\forall q \geq q_0. 
\end{equation}
Then, the error of MPF is bounded by
\begin{equation}\label{Eq_Pre:error_original}
    \norm{e^{-iH\tau} - M_{pmJ}(\tau)} \leq \order{\norm{c}_1 (\mu_{p,m} \tau)^{m+1}},
\end{equation}
where the factor $\mu_{p,m}$ is characterized by the nested commutators as
\begin{equation}\label{Eq_Pre:mu_original}
    \mu_{p,m}\equiv \sup_{\substack{q,n \in \bbN: \\ m+1 \leq q, \\ n \leq \lfloor (q-1)/p  \rfloor}} \left[ \left( \sum_{\substack{p+1\leq q_1,\cdots,q_n: \\ q_1+\cdots+q_n=q+n-1} } \prod_{n'=1}^n \alpha_{\mr{com},q_{n'}} \right)^{\frac{1}{q+n-1}} \right].
\end{equation}
\end{theorem}

Based on this error bound, Ref. \cite{aftab2024-mpf} has evaluated the quantity $\mu_{p,m}$ as $\mu_{p,m} \in \order{N^{\frac1{p+1}} g}$ and obtained that the query complexity in the controlled Trotterization $C[T_p(t/rk_j)]$ amounts to
\begin{equation}\label{Eq_Pre:cost_original}
   \norm{k}_1 \norm{c}_1 r \in \order{N^{\frac1{p+1}}gt \times \mr{polylog} \left( \frac{Ngt}{\varepsilon}\right)}.
\end{equation}
The scaling in the system size $N$ is as large as the one for Trotterization exploiting the commutator scaling, $\order{N^{1/p}}$, and hence this seems to settle the problem on the commutator scaling in MPF.

However, we point out that, although Theorem \ref{Thm:original} itself provides a correct upper bound of the MPF error, it is insufficient to guarantee the favorable size-dependency of the computational cost Eq. (\ref{Eq_Pre:cost_original}) (While we omit the factor $1/q^2$ in Eq. (\ref{Eq_Pre:requirement_original}), it does not change the problem below).
The problem is that the nested commutators $\alpha_{\mr{com},q}$ with infinitely large orders $q$ are involved in the theorem, concretely in Eqs. (\ref{Eq_Pre:requirement_original}) and (\ref{Eq_Pre:mu_original}).
This forces the complexity to obey the $1$-norm scaling, giving no benefit from the locality of the Hamiltonian.
Let us first consider the condition Eq. (\ref{Eq_Pre:requirement_original}).
In order to apply the error bound Eq. (\ref{Eq_Pre:error_original}) to each Trotter step, it is demanded that the time $\tau = t/r$ should satisfy $t/r \leq (\alpha_{\mr{com},q})^{-1/q}$ for arbitrarily large $q$.
If we determine the Trotter number $r$ based on Eq. (\ref{Eq_Pre:commutator_bound}), it should satisfy
\begin{equation}
    r \geq t \left[ (q-1)! (2kg)^{q-1} Ng \right]^{\frac1q}
\end{equation}
for arbitrarily large $q \in \bbN$.
This leads to a meaningless result $r \to \infty$ due to the divergence of $[(q-1)!]^{1/q}$.
On the other hand, we can instead use another bound,
\begin{equation}\label{Eq_Pre:commutator_1_norm}
    \alpha_{\mr{com},q} \leq 2^q \left( \sum_{\gamma=1}^\Gamma \sum_X \norm{h_X^\gamma} \right)^q \in \order{(Ng)^q},
\end{equation}
which gives a meaningful Trotter number called the $1$-norm scaling, $r \in \Omega (Ngt)$.
However, it prohibits the favorable size-dependency like Eq. (\ref{Eq_Pre:cost_original}).
This reflects the fact that the locality plays a central role only in $q$-fold nested commutators with small $q$; while many commutators among disjoint local terms vanish, each term in nested commutators gradually becomes non-local with the spread of the interactions, which provides the factor $q!$ in Eq. (\ref{Eq_Pre:commutator_bound}).
When the order $q$ becomes large enough for the interactions to cover the whole system, such vanishment no longer takes place, and Eq. (\ref{Eq_Pre:commutator_bound}) provides a bound approximately as large as the $1$-norm scaling, Eq. (\ref{Eq_Pre:commutator_1_norm}).

A similar problem takes place in the quantity $\mu_{p,m}$ by Eq. (\ref{Eq_Pre:mu_original}).
When we use the bound Eq. (\ref{Eq_Pre:commutator_bound}) for Eq. (\ref{Eq_Pre:mu_original}), the quantity $\mu_{p,m}$ is bounded by
\begin{equation}
    \mu_{p,m} \leq 2kg \sup_{\substack{q,n \in \bbN: \\ m+1 \leq q, \\ n \leq \lfloor (q-1)/p  \rfloor}} \left\{ \left[ \sum_{\substack{p+1\leq q_1,\cdots,q_n: \\ q_1+\cdots+q_n=q+n-1} } (q_1-1)!\cdots (q_n-1)! (Ng)^n \right]^{\frac{1}{q+n-1}} \right\}.
\end{equation}
The supremum in the above equation is divergent since it should be at least larger than the case with $n=1$ as follows,
\begin{eqnarray}
    && \sup_{q \geq \max(m+1,p+1)} \left\{ \left[ (q-1)! Ng \right]^{\frac{1}{q}} \right\} \nonumber \\
    && \qquad \leq
     \sup_{\substack{q,n \in \bbN: \\ m+1 \leq q, \\ n \leq \lfloor (q-1)/p  \rfloor}} \left\{ \left[ \sum_{\substack{p+1\leq q_1,\cdots,q_n: \\ q_1+\cdots+q_n=q+n-1} } (q_1-1)!\cdots (q_n-1)! (Ng)^n \right]^{\frac{1}{q+n-1}} \right\}.
\end{eqnarray}
The left-hand side is divergent due to $[(q-1)!]^{1/q} \to \infty$.
Namely, the bound reflecting the locality by Eq. (\ref{Eq_Pre:commutator_bound}) cannot give any meaningful upper bound on $\mu_{p,m}$, and hence the scaling $\mu_{p,m} \in \order{N^{1/(p+1)}g}$ predicted by Ref. \cite{aftab2024-mpf} is not concluded. 
Although the bound with the $1$-norm scaling by Eq. (\ref{Eq_Pre:commutator_1_norm}) can provide a meaningful upper bound $\mu_{p,m} \in \order{Ng}$, it prevents the error bound Eq. (\ref{Eq_Pre:error_original}) from concluding the scaling Eq. (\ref{Eq_Pre:cost_original}) as well as the requirement Eq. (\ref{Eq_Pre:requirement_original}).
In Ref. \cite{aftab2024-mpf}, the factor $(q-1)!$ in Eq. (\ref{Eq_Pre:commutator_bound}) is missing since it is often omitted from the scaling in Trotterization with the order $q \in \order{1}$.
This mistakenly leads to the conclusion that Theorem \ref{Thm:original} implies the complexity Eq. (\ref{Eq_Pre:cost_original}).
Therefore, it is still an open problem whether the MPF can achieve the favorable size-dependency like Eq. (\ref{Eq_Pre:cost_original}), i.e., whether the MPF inherits the advantage of the commutator scaling in Trotterization.

The above problem in Ref. \cite{aftab2024-mpf} arises from the necessity of convergence of the BCH formula, $e^A e^B = e^{A+B+[A,B]+\cdots}$.
It allows us to easily compute the powers of Trotterization included in the MPF as
\begin{equation}
    \{ T_p(\tau/k_j) \}^{k_j} = \exp \left( -i H\tau - i \sum_{q=2}^\infty \Phi_q \frac{\tau^q}{(k_j)^{q-1}}\right),
\end{equation}
where an operator $\Phi_q$ is composed of the nested commutators [See Eq. (\ref{Eq_Pf:BCH}) for the explicit formula].
The upper bound on $\norm{\Phi_q}$ is associated with $\alpha_{\mr{com},q}$ by $\norm{\Phi_q} \leq (c_p)^q \alpha_{\mr{com},q}/q^2$ \cite{aftab2024-mpf}, and this is why the error bound includes the nested commutators with sufficiently large $q$.
The BCH expansion should converge for the use in the calculation even when we use its truncation in general. 
This leads to the problematic requirement on the time, Eq. (\ref{Eq_Pre:requirement_original}), which corresponds to the convergence radius $\order{(Ng)^{-1}}$.

\subsection{Summary of our results}\label{Subsec:Summary}

In this section, we summarize our results properly giving the complexity of MPF exploiting the commutator scaling, and highlight the change from the existing analysis \cite{aftab2024-mpf}.
We first prove the following error bound .

\begin{theorem}\label{Thm:ours_informal}
\textbf{(Our error bound of MPF, informal)}

When the time $\tau$ is small enough to satisfy
\begin{equation}\label{Eq_Pre:requirement_ours}
  \tau \in \order{\frac{1}{[N^{\frac1{p+1}}+\log (N/\epsilon)]g}}
\end{equation}
for an arbitrary fixed value $\epsilon \in (0,1)$, the error of MPF is bounded by 
\begin{equation}\label{Eq_Pre:error_ours}
  \norm{e^{-iH\tau}-M_{pmJ}(\tau)} \in \order{\norm{c}_1 (\mu_{p,m}[p_0(N,\epsilon)]\tau)^{m+1}+\norm{c}_1 \norm{k}_1 \epsilon}.
\end{equation}
The quantity $\mu_{p,m}[p_0(N,\epsilon)]$ is defined by
\begin{eqnarray}
    \mu_{p,m}[p_0(N,\epsilon)] &\equiv& \sup_{\substack{q,n \in \bbN: \\ m+1 \leq q, \\ n \leq \lfloor (q-1)/p  \rfloor}} \left[ \left( \sum_{\substack{p+1\leq q_1,\cdots,q_n \textcolor{red}{\leq p_0(N,\epsilon)}: \\ q_1+\cdots+q_n=q+n-1} } \prod_{n'=1}^n \alpha_{\mr{com},q_{n'}} \right)^{\frac{1}{q+n-1}} \right] \label{Eq_Pre:mu_ours} \\
    &\in& \order{[N^{\frac1{p+1}} + \log (N/\epsilon)] g}, \label{Eq_Pre:mu_scaling_ours}
\end{eqnarray}
where $p_0(N,\epsilon) = \lceil \log (3N/\epsilon) \rceil$ denotes the truncation order.

\end{theorem}

There are two significant differences from Theorem \ref{Thm:original}.
The first one is the looseness of the requirement on the time $\tau$.
The time $\tau$ characterized by Eq. (\ref{Eq_Pre:requirement_ours}) can be outside of the convergence radius $\order{(Ng)^{-1}}$.
Although the BCH formula seems to be no longer available, we resolve this problem by the asymptotic convergence under the locality, which was observed in Floquet-Magnus expansions \cite{Kuwahara2016-yn}.
While the requirement Eq. (\ref{Eq_Pre:requirement_original}) forces each Trotter step $t/r$ to be the one corresponding to the $1$-norm scaling, Eq. (\ref{Eq_Pre:requirement_ours}) allows it to be the one for the commutator scaling.
The second difference is the truncation order $p_0(N,\epsilon) \in \order{\log (N/\epsilon)}$ for the nested commutators involved in the error bound, Eqs. (\ref{Eq_Pre:error_ours}) and (\ref{Eq_Pre:mu_ours}).
This prevents the divergent behavior of the upper bound by Eq. (\ref{Eq_Pre:commutator_bound}), which properly reflects the locality of the Hamiltonian, and gives the meaningful bound on the quantity $\mu_{p,m}[p_0(N,\epsilon)]$, Eq. (\ref{Eq_Pre:mu_scaling_ours}).

Based on Theorem \ref{Thm:ours_informal}, we derive the computational cost of Hamiltonian simulation via MPF, which properly reflects the commutator scaling error.
We use the error bound Eq. (\ref{Eq_Pre:error_ours}) with setting $\epsilon \in \order{\varepsilon / (\norm{c}_1 \norm{k}_1 r)}$ to achieve the allowable error $\varepsilon$.
When there is a set of well-conditioned solution for $\{c_j\}$ and $\{k_j \}$ satisfying Eq. (\ref{Eq_Pre:well_condition}), we prove that query complexity in the controlled Trotterization $C[T_p(t/rk_j)]$ amounts to
\begin{equation}
   \norm{k}_1 \norm{c}_1 r \in \order{ \left[ N^{\frac1{p+1}} + (\log (Ngt/\varepsilon))^2 \right] gt \times \mr{polylog} \left( \frac{Ngt}{\varepsilon}\right)}.
\end{equation}
It differs from Eq. (\ref{Eq_Pre:cost_original}), predicted by Ref. \cite{aftab2024-mpf}, by the polylogarithmic factors in $Ngt/\varepsilon$ due to the truncation order $p_0(N,\epsilon)$.
Anyway, our error bound in Theorem \ref{Thm:ours_informal} ensures that the cost of the MPF has a favorable size-dependency by the commutator scaling like Trotterization and also has polylogarithmic dependency in accuracy like LCU and QSVT \cite{Low2017-qsp,Low2019-qubitization,Gilyen2019-qsvt} (See Table \ref{Tab:Gate_counts}).

\section{Modified error bound and complexity of MPF}

In this section, we provide the modified theorems for the commutator scaling in MPF, which is valid for predicting better size-dependency of the error bound and the complexity.
The problem of the original proof \cite{aftab2024-mpf} comes from the fact that the infinite series in the BCH formula should be convergent.
Its convergence radius inversely proportional to the norm of the exponents demands that the time $\tau$ should satisfy $\tau \in \order{(Ng)^{-1}}$. Then, the cost inevitably obeys the $1$-norm scaling \cite{Blanes2009-op-magnus,Watson2024-mpf}.
We expect that the convergence radius cannot generally be improved even in the presence of the locality since the divergence is associated with thermalization to trivial states and it is observed for generic nonintegrable time-periodic systems \cite{Lazarides2014-iv-eth,DAlessio2014-iu-eth,Kuwahara2016-yn}.
The convergence of the BCH formula seems to be incompatible with the good size-dependency reflecting the commutator scaling.

Our strategy is to derive the error bound and the cost without relying on the convergence of the BCH formula.
We show that, under the condition much looser than the convergence radius, $\tau \in \order{(g \log (N/\epsilon))^{-1}}$, Trotterization can be approximated by the truncated BCH formula up to the order $p_0 \in \order{\log (N/\epsilon)}$ within an arbitrarily small error $\epsilon \in (0,1)$.
This is reminiscent of the asymptotic convergence of the Floquet-Magnus expansion for predicting prethermalization under time-periodic Hamiltonians \cite{Abanin2015-zg,Kuwahara2016-yn,Mori2016-tp,Abanin2017-zs,Abanin2017-li,Sharma2024-om-magnus}.
Owing to the truncation order $p_0$, the divergence of the factor $q!$ in Eq. (\ref{Eq_Pre:commutator_bound}) can be circumvented. 
This leads to a meaningful commutator-scaling error of MPF, in which it is sufficient to consider $q$ such that $p<q<p_0$ for Eqs. (\ref{Eq_Pre:error_ours}) and (\ref{Eq_Pre:mu_ours}).
As a result, we clarify that the better size-dependency owing to the commutator-scaling is present also in MPF, while the exact scaling deviates from the one predicted by Ref. \cite{aftab2024-mpf} by a logarithmic quantity.

In Section \ref{Subsec:bch_asymptotic}, we provide the correct commutator-scaling error of MPF.
While we leave some technical parts of the proof in Appendices \ref{SubsecA:bounds} and \ref{SubsecA:Series_expansion}, we show that the nested commutators up to a certain truncation order for the BCH formula are relevant to the MPF error.
In Section \ref{Subsec:cost_MPF}, we derive the Trotter step of MPF for generic local Hamiltonians, which correctly reflects the commutator scaling.

\subsection{Asymptotic convergence of the BCH formula for Trotterization}\label{Subsec:bch_asymptotic}

Here, we prove the asymptotic convergence of the BCH formula for Trotterization, which resolves the strong requirement in Theorem \ref{Thm:original}.
The BCH formula is an infinite series that gives Trotterization $T_p(\tau)$ by
\begin{equation}
    T_p(\tau) = \exp \left( -i H\tau - i \sum_{q=2}^\infty \Phi_q \tau^q \right)
\end{equation}
for sufficiently small time $\tau$ within the convergence radius.
The operator $\Phi_q$ is defined by
\begin{eqnarray}
    && \Phi_q = \sum_{\substack{q_1,\cdots,q_{c_p \Gamma} \geq 0 \\ q_1+\cdots+q_{c_p \Gamma} = q }} (-i)^{q-1} \left( \prod_{v=1}^{c_p \Gamma} \frac{(\alpha_v)^{q_v}}{q_v!}\right) \nonumber \\
    && \qquad \qquad \qquad \qquad\qquad \times \phi_q ( \underbrace{H_1, \cdots, H_1}_{q_1}, \underbrace{H_2, \cdots, H_2}_{q_2}, \cdots, \underbrace{H_{c_p\Gamma}, \cdots, H_{c_p\Gamma}}_{q_{c_p \Gamma}}), \label{Eq_Pf:BCH} \\
    && \phi_q(H_1,\cdots,H_q) = \frac1{q^2} \sum_{\sigma \in S_q} \frac{(-1)^{d_\sigma}}{\left( \begin{array}{c}
         q-1 \\
         d_\sigma
    \end{array}\right)} [H_{\sigma(1)},[H_{\sigma(2)},\cdots,[H_{\sigma(q-1)},H_{\sigma(q)}]]], \label{Eq_Pf:BCH_phi}
\end{eqnarray}
where $S_q$ means the set of permutations of $\{1,\cdots,q\}$ and the number $d_\sigma$ denotes the number of adjacent pairs $(\sigma(i),\sigma(i+1))$ such that $\sigma(i)>\sigma(i+1)$ \cite{arnal2020-bch}.
Each coefficient $\Phi_q$ is equal to $0$ for $q=2,\cdots,p$ from the order condition $T_p(\tau)=e^{-iH\tau}+\order{\tau^{p+1}}$, and bounded by
\begin{equation}\label{Eq_Pf:Phi_q_bound}
    \norm{\Phi_q} \leq \frac{(c_p)^q}{q^2} \alpha_{\mr{com},q}
\end{equation}
for any $q \geq p+1$ \cite{Watson2024-mpf}.
The BCH formula allows us to easily compute $[ T_p(\tau/k_j) ]^{k_j}$ in the MPF.
However, its convergence demands a strong limitation $\norm{\Phi_q} \tau^q < 1$ for arbitrarily large $q$, which results in Eq. (\ref{Eq_Pre:requirement_original}).
As discussed in Section \ref{Sec:Preliminary}, Eq. (\ref{Eq_Pre:commutator_bound}) is not useful for ensuring the convergence due to the factor $q!$.
In general, $\tau \in \order{(Ng)^{-1}}$ from Eq. (\ref{Eq_Pre:commutator_1_norm}) is a known sufficient condition for the convergence, which results in the $1$-norm scaling instead of the  commutator scaling.

To resolve this problem, we prove the following theorem on the asymptotic convergence of the BCH formula for Trotterization.
It states that the BCH formula truncated at a certain order is enough to approximate $T_p(\tau)$ with arbitrarily small error under the much looser requirement on time, Eq. (\ref{Eq_Pre:requirement_ours}).
As opposed to the intuition that the BCH formula is no longer valid for the time out of the convergence radius $t \in \order{(Ng)^{-1}}$, the locality enables us to provide a convergent behavior up to the truncation order as follows.

\begin{theorem}\label{Thm:error_truncated_BCH}
\textbf{(Asymptotic convergence of the BCH formula for Trotterization)}

We set the truncation order of the BCH formula by
\begin{equation}\label{Eq_Pr:truncation_order}
    p_0(N,\epsilon) = \lceil \log (3N/\epsilon) \rceil \in \order{\log (N/\epsilon)}
\end{equation}
for a given arbitrary value $\epsilon \in (0,1)$.
Suppose that the time $\tau>0$ is small enough to satisfy
\begin{equation}\label{Eq_Pf:time_condition}
    |\tau| \leq \frac{1}{8e^3 c_p p_0(N,\epsilon) kg} \in \order{\frac{1}{g \log (N/\epsilon)}} .
\end{equation}
Then, the truncated BCH formula approximates Trotterization $T_p(\tau)$ by
\begin{equation}\label{Eq_Pf:error_truncated_BCH}
    \norm{T_p(\tau)-\exp \left( -iH\tau - i \sum_{q = 2}^{p_0(N,\epsilon)} \Phi_q \tau^q \right)} \leq \epsilon.  
\end{equation}
\end{theorem}

\textbf{Proof.---}
This proof is parallel to the one for the asymptotic convergence of the Floquet-Magnus expansion of time-periodic Hamiltonians \cite{Kuwahara2016-yn}.
For the Hamiltonian term $H_\gamma=\sum_{X \subset \Lambda} h_X^\gamma$ on the lattice $\Lambda = \{1,\cdots,N\}$, we define subsystem Hamiltonians by
\begin{equation}\label{Eq_Pf:subsys_H_def}
    H_\gamma^{> i} = \sum_{j>i} \sum_{\substack{X: X \ni j, \\ X \cap \{1,\cdots,j-1\} = \phi}} h_X^\gamma \
\end{equation}
for each site index $i \in \{0\} \cup \Lambda$.
Each Hamiltonian $H_\gamma^{>i}$ means the collection of interaction terms acting on the sites $\{i+1,\cdots, N\}$, and hence we have $H_\gamma^{>0} = H_\gamma$ and $H_\gamma^{>N} = 0$.
We also define the subsystem Trotterization by
\begin{equation}\label{Eq_Pf:subsys_Trot_def}
    T_p^{>i}(\tau) = \prod_{v=1,\cdots,c_p \Gamma}^\leftarrow e^{-i H_{\gamma_v}^{>i} \alpha_v \tau},
\end{equation}
which corresponds to a collection of operators acting on the sites $\{i+1,\cdots,N\}$ in $T_p(\tau)$ [See Eq. (\ref{Eq_Pre:Trotterization})].
Trotterization $T_p(\tau)$ is trivially decomposed into
\begin{equation}\label{Eq_Pf:decompose_Tp}
    T_p(\tau) = T^{>1}_p(\tau) \left( T^{>1}_p(\tau)^\dagger T^{>0}_p(\tau) \right) = \cdots = \prod_{i=1,\cdots,N}^\leftarrow \left( T^{>i}_p(\tau)^\dagger T^{>i-1}_p(\tau) \right),
\end{equation}
where we used the relation $T^{>N}_p(\tau) = I$. 
We organize a similar decomposition for the truncated BCH formula.
Using a certain truncation order $p_0 \in \bbN$, we define the operators
\begin{equation}\label{Eq_Pf:subsys_BCH_def}
    \tilde{T}_{p,p_0}^{>i}(\tau) = \exp \left( -i H^{>i} \tau - i \sum_{q=2}^{p_0} \Phi_q^{>i} \tau^q \right),
\end{equation}
where $\Phi_q^{>i}$ is obtained by replacing $H_\gamma$ with $H_\gamma^{>i}$ in $\Phi_q$ [See Eq. (\ref{Eq_Pf:BCH})].
It satisfies the order condition $T^{>i}_p(\tau)=\tilde{T}_{p,p_0}^{>i}(\tau)+\order{\tau^{p_0+1}}$.
We decompose the truncated BCH formula by
\begin{equation}\label{Eq_Pf:decompose_BCH}
    \exp \left( -i H \tau - i \sum_{q=2}^{p_0} \Phi_q \tau^q \right) =  \tilde{T}_{p,p_0}^{>0}(\tau) = \prod_{i=1,\cdots,N}^\leftarrow \left(  \tilde{T}_{p,p_0}^{>i}(\tau)^\dagger \tilde{T}_{p,p_0}^{>i-1}(\tau) \right).
\end{equation}
in a similar manner to Eq. (\ref{Eq_Pf:decompose_Tp}).

Using the triangular inequality, the error of interest is bounded by
\begin{equation}
    \norm{T_p(\tau) - \exp \left( -i H \tau - i \sum_{q=2}^{p_0} \Phi_q \tau^q \right)} \leq \sum_{i=1}^N \norm{ T^{>i}_p(\tau)^\dagger T^{>i-1}_p(\tau) - \tilde{T}_{p,p_0}^{>i}(\tau)^\dagger \tilde{T}_{p,p_0}^{>i-1}(\tau)}
\end{equation}
based on the decompositions, Eqs. (\ref{Eq_Pf:decompose_Tp}) and (\ref{Eq_Pf:decompose_BCH}).
Each of the operators $T^{>i}_p(\tau)^\dagger$, $T^{>i-1}_p(\tau)$, $\tilde{T}_{p,p_0}^{>i}(\tau)^\dagger$, and  $\tilde{T}_{p,p_0}^{>i-1}(\tau)$ is analytic in time $\tau$, and we can compose of the series expansions,
\begin{equation}\label{Eq_Pf:subsys_expansion}
    T^{>i}_p(\tau)^\dagger T^{>i-1}_p(\tau) = \sum_{q=0}^\infty T_q^i \tau^q, \quad \tilde{T}_{p,p_0}^{>i}(\tau)^\dagger \tilde{T}_{p,p_0}^{>i-1}(\tau) = \sum_{q=0}^\infty \tilde{T}_q^i \tau^q.
\end{equation}
Owing to the order condition, $T^{>i}_p(\tau)=\tilde{T}_{p,p_0}^{>i}(\tau)+\order{\tau^{p_0+1}}$, we have
\begin{equation}\label{Eq_Pf:error_BCH_triangle}
    \norm{T_p(\tau) - \exp \left( -i H \tau - i \sum_{q=2}^{p_0} \Phi_q \tau^q \right)} \leq \sum_{i=1}^N \sum_{q=p_0+1}^\infty (\|T_q^i\|+\| \tilde{T}_q^i \|) \tau^q.
\end{equation}
Therefore, it is sufficient to obtain the bounds on the coefficients $\norm{T_q^i}$ and $\| \tilde{T}_q^i \|$.
We prove the following inequalities,
\begin{equation}\label{Eq_Pf:series_expansion_result}
    \norm{T_q^i} \leq (4c_p kg)^q, \quad \| \tilde{T}_q^i \| \leq \frac{e^2}2 (8e^2 c_p p_0 kg)^q,
\end{equation}
respectively as Lemmas \ref{LemmaA:Series_subsys_Trot} and \ref{LemmaA:Series_subsys_BCH} in Appendix \ref{SubsecA:Series_expansion}.
We describe the proof in the same appendix due to its technicality.
These bounds come from the $k$-locality and $g$-extensiveness of the Hamiltonian, which is intuitively explained as follows.
Since $T^{>i-1}_p(\tau)$ differs from $T^{>i}_p(\tau)$ only around the site $i$, many terms distant from it cancel with one another in the operator $T^{>i}_p(\tau)^\dagger T^{>i-1}_p(\tau)$.
It effectively forms the structure of causal cones and evolves only the sites around $i$ with the energy scale $\order{g}$.
This is why the order-$q$ coefficient $\norm{T_q^i}$ is proportional to $g^q$ independent of the size $N$ as Eq. (\ref{Eq_Pf:series_expansion_result}).
Similar discussion goes also for the operator $\tilde{T}_{p,p_0}^{>i}(\tau)^\dagger \tilde{T}_{p,p_0}^{>i-1} (\tau)$.
The effective Hamiltonian for the truncated BCH formula, $H + \sum_{q=2}^{p_0} \Phi_q \tau^{q-1}$, is still a local and extensive operator, while the nested commutators there increase the locality and extensiveness (See Lemma \ref{LemmaA:locality_extensive_BCH} in Appendix \ref{SubsecA:bounds}).
The operator $\tilde{T}_{p,p_0}^{>i}(\tau)^\dagger \tilde{T}_{p,p_0}^{>i-1} (\tau)$ has nontrivial actions around the site $i$ as well, which results in $\| \tilde{T}_q^i \| \in \order{(p_0 g)^q}$.
The additional factor $(p_0)^q$ reflects the increase in the locality and extensiveness via the nested commutators.
We consider the small time $\tau$ satisfying Eq. (\ref{Eq_Pf:time_condition}) with setting the truncation order $p_0 = p_0(N,\epsilon)$ as Eq. (\ref{Eq_Pr:truncation_order}).
Then, the inequality Eq. (\ref{Eq_Pf:error_BCH_triangle}) implies
\begin{eqnarray}
    && \norm{T_p(\tau) - \exp \left( -i H \tau - i \sum_{q=2}^{p_0(N,\epsilon)} \Phi_q \tau^q \right)} \nonumber \\
    && \qquad \leq  N \sum_{q=p_0(N,\epsilon)+1}^\infty \left( (4c_p kg\tau)^q + \frac{e^2}2 (8e^2 c_p p_0(N,\epsilon) kg\tau)^q \right) \nonumber \\
    && \qquad \leq N \sum_{q=p_0(N,\epsilon)+1}^\infty \left( e^{-q} + \frac12 e^{-q+2} \right) \nonumber \\ 
    && \qquad \leq 3N e^{-p_0(N,\epsilon)} \leq \epsilon. \qquad \square
\end{eqnarray}

Theorem \ref{Thm:error_truncated_BCH} reflects the asymptotic convergence of Trotterization up to the intermediate order $p_0(N,\epsilon)$.
The time $\tau$ by Eq. (\ref{Eq_Pf:time_condition}) can be out of the convergence radius $\order{(Ng)^{-1}}$.
On the other hand, the order-$q$ coefficient of the BCH formula is bounded by
\begin{equation}
    \norm{\Phi_q \tau^q} \leq \frac{\alpha_{\mr{com},q}\tau^q}{q^2} \leq (2qkg\tau)^q N,
\end{equation}
where we used the bound Eq. (\ref{Eq_Pre:commutator_bound}).
It exponentially decays at least up to $q \in \order{p_0}$ owing to $2qkg \tau<1$.
When the order $q$ further grows from $\order{p_0}$, each term can be large, which leads to the divergence of the BCH formula.
Theorem \ref{Thm:error_truncated_BCH} states that the truncated BCH formula convergent up to the order $p_0(N,\epsilon)$ can approximate Trotterization $T_p(\tau)$ with a desirable error $\epsilon$.

While we borrow the proof idea from the Floquet-Magnus expansion for generic time-periodic Hamiltonians \cite{Kuwahara2016-yn}, we note that Theorem \ref{Thm:error_truncated_BCH} gives a better bound specifically reflecting the form of Trotterization and the BCH formula via Lemmas \ref{LemmaA:Series_subsys_Trot} and \ref{LemmaA:Series_subsys_BCH}.
This improvement is essential for the present error analysis to be available to MPF for generic local Hamiltonians including long-ranged interacting cases (See Appendix \ref{SubsecA:improvement} for the details).

\subsection{Error bound of MPF with commutator scaling}

Next, we prove the error bound of MPF, Theorem \ref{Thm:ours_informal}.
The proof is done in the following two steps based on the truncated BCH formula in Theorem \ref{Thm:error_truncated_BCH}.
First, we derive the error bound expressed by the nested commutators up to the truncation order $p_0(N,\epsilon)$ in Theorem \ref{Thm:error_MPF}.
Next, we show the scaling of the quantity $\mu_{p,m}[p_0(N,\epsilon)]$, which characterizes the error bound by Eq. (\ref{Eq_Pre:mu_ours}), in Lemma \ref{Lemma:mu_bound}.
The first result is described as follows.

\begin{theorem}\label{Thm:error_MPF}
\textbf{}

For a $k$-local and $g$-extensive Hamiltonian $H$, we define a value,
\begin{equation}\label{Eq_Pr:mu_truncated}
    \mu_{p,m}[p_0] \equiv \sup_{\substack{q,n \in \bbN: \\ m+1 \leq q, \\ n \leq \lfloor (q-1)/p \rfloor}} \left[\left( \sum_{\substack{p+1\leq q_1,\cdots,q_n \leq p_0 \\ q_1+\cdots+q_n=q+n-1} } \prod_{n'=1}^n \alpha_{\mr{com},q_{n'}} \right)^{\frac{1}{q+n-1}} \right],
\end{equation}
which is determined by the nested commutators $\alpha_{\mr{com},q}$ by Eq. (\ref{Eq_Pre:commutator_bound}).
When the time $\tau$ is small enough to satisfy
\begin{equation}\label{Eq_Pf:time_condition_error_MPF}
    |\tau| \leq \min \left( \frac{1}{8e^3 c_p p_0(N,\epsilon)kg}, \frac{1}{2 c_p \mu_{p,m}[p_0(N,\epsilon)]}\right),
\end{equation}
for an arbitrary value $\epsilon \in (0,1)$, the error of MPF is bounded by
\begin{equation}\label{Eq_Pf:error_MPF}
    \norm{e^{-iH\tau} - M_{pmJ}(\tau)} \leq 2 e^{1/2} \norm{c}_1 \left\{ c_p \mu_{p,m}[p_0(N,\epsilon)] \tau\right\}^{m+1} + \norm{c}_1 \norm{k}_1 \epsilon.
\end{equation}
The truncation number $p_0(N,\epsilon)$ is given by Eq. (\ref{Eq_Pr:truncation_order}).

\end{theorem}

\textbf{Proof.---} We use the relation,
\begin{equation}\label{Eq_Pf:Dyson}
    e^{A+B} = e^A \sum_{n=0}^\infty \int_0^1 \dd s_1 \cdots \int_0^{s_{n-1}} \dd s_n \prod_{n'=1,\cdots,n}^\rightarrow \left(e^{-A s_{n'}} B e^{As_{n'}} \right),
\end{equation}
which is equivalent to the Dyson series expansion in the interaction picture based on the operator $A$.
With setting $A=-iH\tau$ and $B=-i \sum_{q=2}^{p_0} \Phi_q \tau^q/(k_j)^{q-1}$, we obtain
\begin{eqnarray}
    && \exp \left( -iH\tau - i \sum_{q = 2}^{p_0(N,\epsilon)} \Phi_q \frac{\tau^q}{(k_j)^{q-1}} \right) - e^{-iH\tau} \nonumber \\
    && \quad = e^{-iH\tau} \sum_{n=1}^\infty \int_0^1 \dd s_1 \cdots \int_0^{s_{n-1}} \dd s_n \prod_{n'=1,\cdots,n}^\rightarrow \left( -i e^{is_{n'} \tau \ad_H} \sum_{q=2}^{p_0(N,\epsilon)} \Phi_q \frac{\tau^q}{(k_j)^{q-1}}\right) \nonumber \\
    && \quad =  e^{-iH\tau}\sum_{n=1}^\infty \sum_{q=pn+1}^{p_0(N,\epsilon)n+1} \frac{\tau^{q+n-1}}{(k_j)^{q-1}} \nonumber \\
    && \qquad \qquad \qquad \qquad  \times \sum_{\substack{p+1 \leq q_1,\cdots,q_n \leq p_0(N,\epsilon) \\ q_1+\cdots+q_n=q+n-1}} \int_0^1 \dd s_1 \cdots \int_0^{s_{n-1}} \dd s_n \prod_{n'=1,\cdots,n}^\rightarrow \left( -i e^{is_{n'} \tau \ad_H} \Phi_{q_{n'}}\right) \nonumber \\
    && \quad =  e^{-iH\tau} \sum_{q=p+1}^\infty  \frac{\tau^q}{(k_j)^{q-1}} \sum_{n=\lceil \frac{q-1}{p_0(N,\epsilon)}\rceil}^{\lfloor \frac{q-1}p \rfloor} \tau^{n-1}  \nonumber \\
    && \qquad \qquad \times \sum_{\substack{p+1 \leq q_1,\cdots,q_n \leq p_0(N,\epsilon) \\ q_1+\cdots+q_n=q+n-1}} \int_0^1 \dd s_1 \cdots \int_0^{s_{n-1}} \dd s_n \prod_{n'=1,\cdots,n}^\rightarrow \left( -i e^{is_{n'} \tau \ad_H} \Phi_{q_{n'}}\right).
\end{eqnarray}
We use $\Phi_q=0$ for $q=2,\cdots,p$ from the order condition $T_p(\tau)=e^{-iH\tau}+\order{\tau^{p+1}}$ in the second inequality.
When we choose $\{ c_j \}$ and $\{ k_j \}$ satisfying the order condition Eq. (\ref{Eq_Pre:MPF_order}), the terms with $q \leq m$ vanish.
As a result, we arrive at
\begin{eqnarray}
    && \norm{\sum_{j=1}^J c_j \exp \left(-iH\tau - i \sum_{q = 2}^{p_0(N,\epsilon)}  \frac{\Phi_q \tau^q}{(k_j)^{q-1}}\right) - e^{-iH\tau}} \nonumber \\
    && \qquad \leq \sum_{j=1}^J |c_j| \sum_{q=m+1}^\infty \sum_{n=1}^{\lfloor \frac{q-1}{p} \rfloor} \frac{\tau^{q+n-1}}{(k_j)^{q-1}} \sum_{\substack{p+1 \leq q_1,\cdots,q_n \leq p_0(N,\epsilon) \\ q_1+\cdots+q_n=q+n-1}} \frac1{n!} \prod_{n'=1}^n \norm{\Phi_{q_{n'}}} \nonumber \\
    && \qquad \leq \norm{c}_1 \sum_{q=m+1}^\infty \sum_{n=1}^\infty \frac{(c_p \mu_{p,m}[p_0(N,\epsilon)] \tau)^{q+n-1}}{n!} \nonumber \\
    && \qquad \leq \norm{c}_1 \sum_{q=m+1}^\infty (c_p \mu_{p,m}[p_0(N,\epsilon)]\tau)^q e^{c_p \mu_{p,m}[p_0(N,\epsilon)]\tau} \nonumber \\
    && \qquad \leq 2 e^{1/2} \norm{c}_1 (c_p\mu_{p,m}[p_0(N,\epsilon)]\tau)^{m+1}.  
\end{eqnarray}
We used the definition of the quantity $\mu_{p,m}[p_0]$ by Eq. (\ref{Eq_Pr:mu_truncated}) and the bound on $\norm{\Phi_q}$ by Eq. (\ref{Eq_Pf:Phi_q_bound}) in the second inequality.
We also used the condition on the time $\tau$ by Eq. (\ref{Eq_Pf:time_condition_error_MPF}) in the third and fourth inequalities.
The error bound of MPF results can be evaluated by the triangle inequality, which results in
\begin{eqnarray}
    \norm{e^{-iH\tau}-M_{pmJ}(\tau)} &\leq& 2 e^{1/2} \norm{c}_1 (c_p\mu_{p,m}[p_0(N,\epsilon)]\tau)^{m+1} \nonumber \\
    && \quad  + \sum_{j=1}^J |c_j| \norm{[T_p(\tau/k_j)]^{k_j} - \exp \left( -iH\tau-i\sum_{q=2}^{p_0(N,\epsilon)} \frac{\Phi_q \tau^q}{(k_j)^{q-1}}\right)} \nonumber \\
    &\leq& 2 e^{1/2} \norm{c}_1 (c_p\mu_{p,m}[p_0(N,\epsilon)]\tau)^{m+1} \nonumber \\
    && \, + \sum_{j=1}^J |c_j| \times |k_j| \norm{T_p(\tau/k_j)-\exp\left(-iH\tau/k_j -i \sum_{q=2}^{p_0(N,\epsilon)} \frac{\Phi_q \tau^q}{(k_j)^q}\right)} \nonumber \\
    && \quad \leq 2 e^{1/2} \norm{c}_1 (c_p\mu_{p,m}[p_0(N,\epsilon)]\tau)^{m+1} + \norm{c}_1 \norm{k}_1 \epsilon,
\end{eqnarray}
where we used the asymptotic convergence of the BCH formula by Theorem \ref{Thm:error_truncated_BCH} in the last inequality. $\quad \square$

The difference from the original statement in Ref. \cite{aftab2024-mpf} (i.e., Theorem \ref{Thm:original}) is the presence of the truncation order $p_0(N,\epsilon) \in \order{\log(N/\epsilon)}$ in Eq. (\ref{Eq_Pr:mu_truncated}).
This brings a meaningful bound reflecting the commutator scaling on the quantity $\mu_{p,m}[p_0(N,\epsilon)]$.

\begin{lemma}\label{Lemma:mu_bound}
\textbf{}

The quantity $\mu_{p,m}[p_0(N,\epsilon)]$, which is defined by Eq. (\ref{Eq_Pr:mu_truncated}), is bounded by
\begin{equation}\label{Eq_Pf:mu_bound}
  \mu_{p,m}[p_0(N,\epsilon)] \leq  4 \max \left[ (p+1) N^{\frac1{p+1}} , e^3 p_0(N,\epsilon) \right] kg\in \order{\left[N^{\frac1{p+1}}+\log (N/\epsilon) \right] g},
\end{equation}
for a value $\epsilon \in (0,1)$.
\end{lemma}

\textbf{Proof.---}
Using the upper bound on the nested commutator by Eq. (\ref{Eq_Pre:commutator_bound}), we obtain
\begin{eqnarray}
  \sum_{\substack{p+1\leq q_1,\cdots,q_n \leq p_0(N,\epsilon) \\ q_1+\cdots+q_n=q+n-1} } \prod_{n'=1}^n \alpha_{\mr{com},q_{n'}} &\leq& \sum_{\substack{p+1\leq q_1,\cdots,q_n \leq p_0(N,\epsilon) \\ q_1+\cdots+q_n=q+n-1} } \prod_{n'=1}^n \left[ (q_{n'}-1)! (2kg)^{q_{n'}-1} Ng \right] \nonumber \\
  &\leq& \sum_{\substack{p+1\leq q_1,\cdots,q_n \leq p_0(N,\epsilon) \\ q_1+\cdots+q_n=q+n-1} } 2^{-n} \prod_{n'=1}^n (2q_{n'} N^{1/q_{n'}}kg)^{q_{n'}}. \label{Eq_Pf:mu_truncated_prf1}
\end{eqnarray}
The function $x N^{1/x}$ is monotonically decreasing in $0<x<\log N$ and increasing in $\log N < x$, which results in
\begin{eqnarray}
  \max_{p+1 \leq q_{n'} \leq p_0(N,\epsilon)} \left( q_{n'} N^{1/q_{n'}}\right) &\leq& \max \left[ (p+1) N^\frac{1}{p+1}, p_0(N,\epsilon) N^{\frac1{p_0(N,\epsilon)}} \right] \nonumber \\
  &\in& \order{N^\frac{1}{p+1} + p_0(N,\epsilon)}.
\end{eqnarray}
We note that the factor $N^{\frac1{p_0(N,\epsilon)}}$ is bounded by a constant as $N^{\frac1{p_0(N,\epsilon)}} \leq N^{\frac1{\log (3N/\epsilon)}} = e^{1-\frac{\log (3/\epsilon)}{\log N + \log (3/\epsilon)}} < e^3$.
Substituting this result into Eq. (\ref{Eq_Pf:mu_truncated_prf1}), we obtain
\begin{eqnarray}
  && \sum_{\substack{p+1\leq q_1,\cdots,q_n \leq p_0(N,\epsilon) \\ q_1+\cdots+q_n=q+n-1} } \prod_{n'=1}^n \alpha_{\mr{com},q_{n'}} \nonumber \\
  && \qquad \leq 2^{-n} \left\{ \max \left( (p+1) N^\frac{1}{p+1}, e^3 p_0(N,\epsilon) \right) kg \right\}^{q+n-1} \sum_{\substack{q_1,\cdots,q_n \geq 0: \\ q_1+\cdots+q_n=q+n-1}} 1 \nonumber \\
  && \qquad \leq \left[ 4 \max \left( (p+1) N^\frac{1}{p+1}, e^3 p_0(N,\epsilon) \right) kg \right]^{q+n-1}.
\end{eqnarray}
The summation in the right-hand side of the first line is equal to $(q+2n-2)!/(n-1)!(q+n-1)!$, which is bounded by $2^{q+2n-2}$.
This completes the proof. $\quad \square$

Theorem \ref{Thm:ours_informal} immediately follows from the combination of Theorem \ref{Thm:error_MPF} and this lemma.
Using the scaling of $\mu_{p,m}[p_0]$ by Eq. (\ref{Eq_Pf:mu_bound}), the requirement on the time $\tau$ in Theorem \ref{Thm:error_MPF} is satisfied under
\begin{equation}
  |\tau| \in \order{\frac{1}{\left[N^{\frac1{p+1}}+\log (N/\epsilon) \right] g}}.
\end{equation}
Then, the error bound by Eq. (\ref{Eq_Pf:error_MPF}) results in
\begin{equation}
  \norm{e^{-iH\tau}-M_{pmJ}(\tau)} \in \order{\norm{c}_1\left[ (N^{\frac1{p+1}}+\log(N/\epsilon)) gt \right]^{m+1}+ \norm{c}_1\norm{k}_1 \epsilon}.
\end{equation}
As briefly discussed in Section \ref{Subsec:Summary}, our theorem requires a much looser condition on the time $\tau$ and provides a meaningful error bound by commutator scaling compared to the original one in Theorem \ref{Thm:original}.
This comes from the fact that we rely on the asymptotic convergence of the BCH formula rather than the convergence, bringing the truncation order $p_0(N,\epsilon)$ valid even out of the convergence radius.

\subsection{Computational cost of the well-conditioned MPF}\label{Subsec:cost_MPF}

We next derive the complexity of MPF based on Theorem \ref{Thm:ours_informal}, which leads to the complexity inheriting the benefit of commutator scaling.
The main result is described by the following theorem.

\begin{theorem}\label{Thm:trotter_number}
\textbf{(Complexity of MPF)}

Suppose that there exists a well-conditioned solution for $\{c_j\}$ and $\{k_j\}$ satisfying Eq. (\ref{Eq_Pre:well_condition}) and that the order $m$ is proportional to the number of terms $J$.
The query complexity in the controlled Trotterization $C[T_p(\tau)]$ can be
\begin{equation}\label{Eq_Pf:MPF_complexity}
    \order{\left\{ N^{\frac1{p+1}} + \left[\log (N gt/\varepsilon)\right]^2 \right\} gt \times \mr{polylog} (Ngt/\varepsilon)}.
\end{equation}
for simulating $e^{-iHt}$ within an additive error $\varepsilon$ by the MPF $M_{pmJ}(\tau)$.
\end{theorem}

\textbf{Proof.---}
The algorithm based on MPF runs with the original proposal by Ref. \cite{low2019-mpf}.
We split the time $t$ into $r$ parts so that the error at each Trotter step is bounded by
\begin{equation}
  \norm{e^{-iHt/r}-M_{pmJ}(t/r)} \leq \frac{\varepsilon}{2r},
\end{equation}
which is a sufficient condition for Eq. (\ref{Eq_Pre:r_condition}) \cite{aftab2024-mpf}.
We deterministically implement $M_{pmJ}(t/r)$ by the LCU approach combined with QAA.
We repeat this $r$ times, and then we can reproduce $e^{-iHt}$ within an error $\varepsilon$.
The number of queries to the controlled Trotterization $C[T_p(t/(rk_j))]$ amounts to $\order{\norm{c}_1 \norm{k}_1 r}$.

We set the value $\epsilon \in (0,1)$ by $\epsilon = \varepsilon / (4 \norm{c}_1 \norm{k}_1 r)$ in Theorem \ref{Thm:error_MPF}.
According to the error bound by Eq. (\ref{Eq_Pf:error_MPF}) and the quantity $\mu_{p,m}[p_0(N,\epsilon)]$ by Lemma \ref{Lemma:mu_bound}, it is sufficient to find the number $r$ such that both of the following two inequalities are satisfied,
\begin{equation}\label{Eq_Pf:2_ineq_for_r}
    \begin{cases}\displaystyle
    2 \times 2^m \norm{c}_1 \left[ 4 c_p  (p+1)N^{\frac1{p+1}} kg  \frac{t}{r} \right]^{m+1} + \norm{c}_1\norm{k}_1 \frac{\varepsilon}{4\norm{c}_1\norm{k}_1r}\leq \frac{\varepsilon}{2r}, \\
    \displaystyle
    2 \times 2^m \norm{c}_1 \left[ 4 e^3 c_p p_0(N,4\norm{c}_1\norm{k}_1r/\varepsilon) kg  \frac{t}{r} \right]^{m+1} + \norm{c}_1\norm{k}_1 \frac{\varepsilon}{4\norm{c}_1\norm{k}_1r}\leq \frac{\varepsilon}{2r}.
    \end{cases}
\end{equation}
We replace $e^{1/2}$ by $2^m$ ($>e^{1/2}$) in the upper bound for the later purpose to show the satisfaction of the requirement, Eq. (\ref{Eq_Pf:time_condition_error_MPF}).
The first one is easily solved as well  as the ordinary Trotterization, which results in
\begin{equation}\label{Eq_Pf:first_condition_r}
    r \geq 8 c_p  (p+1)k N^{\frac1{p+1}} gt \left[ \frac{32 c_p (p+1)k \norm{c}_1 N^{\frac1{p+1}}gt}{\varepsilon}\right]^{\frac1m} \equiv r_{1,m} (N,t,\varepsilon).
\end{equation}

In the second inequality, we see that the left-hand side involves an additional logarithmic factor $p_0(N,4\norm{c}_1\norm{k}_1r/\varepsilon)$ by Eq. (\ref{Eq_Pr:truncation_order}).
Here, we use the following relation,
\begin{equation}\label{Eq_Pf:log_x_over_x}
  \frac{(\log x +1)^{m+1}}{x^m} \leq a,\quad ^\forall x \geq 5^{1+\frac1m} a^{-\frac1m} \log^{1+\frac1m}(1/a), \quad a \in (0,1/5].
\end{equation}
This can be easily confirmed by the substitution $x_a \equiv 5^{1+\frac1m} a^{-\frac1m} \log^{1+\frac1m}(1/a)$, which leads to
\begin{eqnarray}
    \frac{(\log x_a +1)^{m+1}}{(x_a)^m} &=& \frac{a}{[5 \log (1/a) ]^{m+1}} \left\{ \left( 1+\frac1m \right) \log [5 \log (1/a)] + \frac1m \log (1/a) + 1 \right\}^{m+1} \nonumber \\
    &\leq& a,
\end{eqnarray}
and the monotonicity of $(\log x + 1)^{m+1}/x^m$ in $x>1$.
We used the relations $m \geq 1$, $\log (1/a) \geq \log 5$, and $\frac12 \log (1/a) \geq \log \log (1/a)$ in the above.
Taking into account that the truncation order satisfies
\begin{equation}
    p_0(N,4\norm{c}_1\norm{k}_1r/\varepsilon) \leq \log \left( \frac{12 \norm{c}_1 \norm{k}_1 N}{\varepsilon} r\right) + 1
\end{equation}
based on Eq. (\ref{Eq_Pr:truncation_order}), we exploit the inequality Eq. (\ref{Eq_Pf:log_x_over_x}) with setting the values $x$ and $a$ respectively by 
\begin{equation}
  x = \frac{12 \norm{c}_1 \norm{k}_1 N}{\varepsilon} r, \quad a = \frac{\varepsilon}{4\norm{c}_1(8e^3 c_p kgt)^{m+1}} \left( \frac{\varepsilon}{12\norm{c}_1\norm{k}_1 N} \right)^m.
\end{equation}
Since we are interested in scaling behavior under sufficiently large $N$, $t$, or $1/\varepsilon$ and we have $\norm{c}_1,\norm{k}_1 \geq 1$, the above value $a$ is assumed to be $a \in (0,1/5]$ without loss of generality.
Then, we have
\begin{eqnarray}
    && \frac{\varepsilon}{12\norm{c}_1\norm{k}_1 N} \times 5^{1+\frac1m} a^{-\frac1m} \log^{1+\frac1m}(1/a)  \nonumber \\
    && \qquad = 40 e^3 c_p kgt \left( \frac{160 e^3 \norm{c}_1 c_p kgt}{\varepsilon}\right)^{\frac1m} \nonumber \\
    && \qquad \qquad \times \left\{ \log \left[\frac{(8e^3 c_p kgt)(12 \norm{c}_1\norm{k}_1N)}{\varepsilon}\right]^{m+1} - \log \left( 3\norm{k}_1 N\right)\right\}^{1+\frac1m}.
\end{eqnarray}
Since we have the relation $(m+1)^{1+\frac1m} = (m+1) e^{m^{-1}\log (m+1)} \leq e(m+1)$, the sufficient condition for the second inequality in Eq. (\ref{Eq_Pf:2_ineq_for_r}) is
\begin{eqnarray}
    r &\geq& 40 e^4 c_p kgt (m+1) \left( \frac{160 e^3 \norm{c}_1 c_p kgt}{\varepsilon}\right)^{\frac1m} \log^{1+\frac1m} \left[\frac{(8e^3 c_p kgt)(12 \norm{c}_1\norm{k}_1N)}{\varepsilon}\right] \nonumber \\
    &\equiv& r_{2,m}(N,t,\varepsilon). \label{Eq_Pf:Second_condition_r}
\end{eqnarray}
Thus, we  can set the number $r$ by $r = \lceil \max (r_{1,m}(N,t,\varepsilon),r_{2,m}(N,t,\varepsilon)) \rceil$ [See Eq. (\ref{Eq_Pf:first_condition_r}) for $r_{1,m}(N,t,\varepsilon)$]. 

The order of the MPF $m$ is determined so that the polynomial terms with the exponent $1/m$ can be bounded by constants in both Eqs. (\ref{Eq_Pf:first_condition_r}) and (\ref{Eq_Pf:Second_condition_r}).
For instance, it is sufficient to choose the order $m$ by
\begin{equation}\label{Eq_Pf:r_proof6}
  m = \left\lceil \log \left( \frac{Ngt}{\varepsilon}\right) \right\rceil \in \order{\log \left( \frac{Ngt}{\varepsilon}\right)}.
\end{equation}
When we use the well-conditioned solution by Eq. (\ref{Eq_Pre:well_condition}) and assume the order $m \propto J$, we have
\begin{equation}
    \norm{c}_1 \in \mr{polylog} \left( \frac{Ngt}{\varepsilon}\right), \quad (\norm{c}_1)^{\frac1m} \in e^{\Theta (\log m)/m} \subset \order{1}.
\end{equation}
The quantity $r_{1,m}(N,t,\varepsilon)$ given by Eq. (\ref{Eq_Pf:first_condition_r}) scales as 
\begin{eqnarray}
    r_{1,m}(N,t,\varepsilon) &\in& \Theta \left( N^{\frac1{p+1}}gt (\norm{c}_1)^{\frac1m} \left( \frac{N^{\frac1{p+1}}gt}{\varepsilon}\right)^{\frac1m} \right) \nonumber \\
    &=& \Theta \left( N^{\frac1{p+1}}gt \right). 
\end{eqnarray}
In a similar way, the quantity $r_{2,m}(N,t,\varepsilon)$ given by Eq. (\ref{Eq_Pf:Second_condition_r}) scales as
\begin{eqnarray}
    r_{2,m}(N,t,\varepsilon) &\in& \Theta \left( mgt (\norm{c}_1)^{\frac1m}\left( \frac{gt}{\varepsilon}\right)^{\frac1m} \log^{1+\frac1m} \left( \frac{\norm{c}_1 \norm{k}_1 Ngt}{\varepsilon}\right) \right) \nonumber \\
    &=& \Theta \left( gt \left( \log \frac{Ngt}{\varepsilon}\right)^2 \right),
\end{eqnarray}
where we use $\norm{k}_1 \in \mr{polylog}(Ngt/\varepsilon)$ from the well condition, Eq. (\ref{Eq_Pre:well_condition}).
The number $r$ is at most bounded by
\begin{equation}\label{Eq_Pf:r_scaling}
  r \leq r_{1,m}(N,t,\varepsilon) + r_{2,m}(N,t,\varepsilon) + 1 \in \order{\left[ N^{\frac1{p+1}} + \left( \log \frac{Ngt}{\varepsilon}\right)^2  \right] gt}.
\end{equation}
As discussed at the beginning of the proof, the algorithm running the LCU approach combined with QAA requires $\order{\norm{c}_1 \norm{k}_1 r}$ queries to the operator $C[T_p(t/(rk_j))]$.
Multiplying $\norm{c}_1 \norm{k}_1 \in \mr{polylog}(Ngt/\varepsilon)$ to Eq. (\ref{Eq_Pf:r_scaling}), we arrive at Eq. (\ref{Eq_Pf:MPF_complexity}) as the query complexity.

Finally, the above cost evaluation relies on the error bound by Theorem \ref{Thm:error_MPF}.
We have to confirm that the above situation is compatible with the requirement, Eq. (\ref{Eq_Pf:time_condition_error_MPF}).
Since the number $r$ is determined so that the two inequalities in Eq. (\ref{Eq_Pf:2_ineq_for_r}) hold, each time step $\tau = t/r$ is bounded by
\begin{eqnarray}
    \frac{t}{r} &\leq& \left\{ 8 c_p \max \left[ (p+1) N^{\frac1{p+1}}, e^3 p_0(N,4 \norm{c}_1\norm{k}_1r/\varepsilon) \right] kg \right\}^{-1} \left( \frac{\varepsilon}{4 \norm{c}_1 r}\right)^{\frac1{m+1}} \nonumber \\
    &\leq& \min \left( \frac{1}{8e^3 c_p p_0(N,4 \norm{c}_1\norm{k}_1r/\varepsilon) kg}, \frac{1}{2c_p \mu_{p,m}[p_0(N,4 \norm{c}_1\norm{k}_1r/\varepsilon)]} \right) \nonumber \\
    && \qquad  \qquad \qquad \qquad \qquad  \qquad \qquad \qquad  \qquad \times \exp \left( - \frac{\log (4 \norm{c}_1r/\varepsilon)}{\lceil \log (Ngt/\varepsilon)\rceil + 1}\right). \label{Eq_Pf:requirement_satisfied}
\end{eqnarray}
From Eqs. (\ref{Eq_Pf:r_proof6}) and (\ref{Eq_Pf:r_scaling}), the last exponential term is an $\order{1}$ positive constant smaller than $1$.
Therefore, the requirement Eq. (\ref{Eq_Pf:time_condition_error_MPF}) is satisfied for the time step $\tau=t/r$ and we conclude that the query complexity Eq. (\ref{Eq_Pf:MPF_complexity}) based on Theorem \ref{Thm:error_MPF} is indeed valid. $\quad \square$

Theorem \ref{Thm:trotter_number} establishes the cost of the MPF-based algorithm properly reflecting its commutator-scaling errors.
The previous result by Theorem \ref{Thm:original} is incompatible with the cost benefiting from the commutator scaling as discussed in Section \ref{Subsec:Problem}.
By contrast, our result allows it as we have confirmed by Eq. (\ref{Eq_Pf:requirement_satisfied}).
We summarize the gate counts of various quantum algorithms for Hamiltonian simulation in Table \ref{Tab:Gate_counts}.
The MPF-based algorithm has $\mr{polylog}(1/\varepsilon)$ dependency in its cost like the LCU or QSVT approaches, as proven in Ref. \cite{low2019-mpf}.
Our central result is its size-dependency $\tilde{\mcl{O}}(N^{1+1/(p+1)})$ like Trotterization, which is derived from the error bound, Theorem \ref{Thm:error_MPF}.
We also display the Haah-Hastings-Kothari-Low (HHKL) algorithms \cite{Haah2021-time-dep} for comparison, which achieves the near-optimal gate count $\tilde{\mcl{O}}(Nt)$ for generic Hamiltonians with finite-range interactions.
The MPF has slightly larger size-dependence by the factor $N^{1/(p+1)}$ for finite-ranged systems.
On the other hand, it offers a versatile efficient approach for generic local Hamiltonians including long-ranged interacting cases.
This comes from the fact that the HHKL algorithm heavily relies on the Lieb-Robinson bound \cite{Lieb1972-uo} while the MPF does not.

\begin{table}[]
  \centering
\small
  \begin{tabular}{|c||c|c|}
     & \begin{tabular}{c} Gate counts for Hamiltonians \\ with finite-range interactions \end{tabular}
      & \begin{tabular}{c}
        Gate counts for Hamiltonians \\ with long-range interactions
     \end{tabular} \\ \hline \hline
    \begin{tabular}{c} Trotterization \\ (Ref. \cite{childs2021-trotter}) \end{tabular} & \( \displaystyle Ngt \left(\frac{Ngt}{\varepsilon}\right)^{\frac1p}\) & \( \displaystyle N^k gt \left(\frac{Ngt}{\varepsilon}\right)^{\frac1p}\) \\ \hline
    \begin{tabular}{c} LCU \\ (Ref. \cite{Berry-prl2015-LCU}) \end{tabular} & \(\displaystyle N^2 gt \frac{\log (Ngt/\varepsilon)}{\log \log(Ngt/\varepsilon)}\) &  \(\displaystyle N^{k+1} gt \frac{\log (Ngt/\varepsilon)}{\log \log(Ngt/\varepsilon)}\)  \\ \hline
    \begin{tabular}{c} QSVT \\ (Refs. \cite{Low2019-qubitization,Gilyen2019-qsvt}) \end{tabular}  &  \(\displaystyle N \left[ Ngt +  \frac{\log (1/\varepsilon)}{\log \log(1/\varepsilon)} \right] \) & \(\displaystyle N^k \left[ Ngt +  \frac{\log (1/\varepsilon)}{\log \log(1/\varepsilon)} \right] \)  \\ \hline
    \begin{tabular}{c} MPF \\ (Our result) \end{tabular} & \( \displaystyle N^{1+\frac1{p+1}} gt \, \mr{polylog}(Ngt/\varepsilon)\) & \( \displaystyle N^{k+\frac1{p+1}} gt \, \mr{polylog}(Ngt/\varepsilon)\) \\ \hline
    \begin{tabular}{c} HHKL algorithm \\ (Refs. \cite{Haah2021-time-dep,Tran-PRX2019-hhkl}) \end{tabular} & \( \displaystyle Ngt \, \mr{polylog}(Ngt/\varepsilon)\) & \(\displaystyle Ng t \left( \frac{Ngt}{\varepsilon} \right)^{\frac{2d}{\nu-d}} \) [Only for $\nu > 2d$] \\ \hline
  \end{tabular}
  
  \caption{Gate counts for Hamiltonian simulation, which are measured by the number of $\order{1}$-qubit gates (they can be geometrically non-local). These are computed based on the query complexity multiplied by the gate counts per oracle, where the oracles are $T_p(\tau)$ and $C[T_p(\tau)]$ for some small time $\tau$ respectively for Trotterization and MPF, and those for LCU and QSVT are block-encoding operators of the Hamiltonian. In any case, the oracles typically require $\order{N}$ gates (for finite-ranged interactions) or $\order{N^k}$ gates (for $k$-local, long-ranged interactions), which are proportional to the number of terms in the Hamiltonian. For the MPF, we omit $[\log(Ngt/\varepsilon)]^2$ in the query complexity, Eq. (\ref{Eq_Pf:MPF_complexity}), since it can be absorbed into the polylogarithmic term.}
  \label{Tab:Gate_counts}
\end{table}

\section{Discussion and conclusion}

\subsection{Application to other quantum algorithms}

Recently, there have been various quantum algorithms trying to achieve better cost both in the size and accuracy like MPF \cite{Endo-PRA2019-mitigation,Faehrmann_2022-mpf,Watson2024-mpf,Rendon2023-interpolate,Rendon2024-interpolate,watson2024-mpf-random,chakraborty2025-mpf,Guo2025-interpolate}, such as the sampling-based approach instead of relying on LCU \cite{Faehrmann_2022-mpf}.
In particular, the mitigation of Trotter errors with polynomial interpolation or Richardson extrapolation has attracted much interest \cite{Endo-PRA2019-mitigation}.
They are exemplified by an algorithm for time-evolved observables \cite{Watson2024-mpf} and those for generic tasks including quantum signal processing \cite{Rendon2023-interpolate,Rendon2024-interpolate,chakraborty2025-mpf,Guo2025-interpolate}.
These algorithms are expected to have favorable cost with respect to the system size owing to the commutator scaling. 
Indeed, some of them \cite{Watson2024-mpf,chakraborty2025-mpf} have derived their error and cost expressed by nested commutators with the same strategy as the original proof for MPF \cite{aftab2024-mpf}.
The regimes where the time at each step is small enough for the BCH formula to converge are also assumed in Refs. \cite{Rendon2023-interpolate,Rendon2024-interpolate,Guo2025-interpolate}.
As discussed in Section \ref{Subsec:Problem}, error bounds expressed by nested commutators do not necessarily imply the cost reflecting the commutator scaling in such regimes. 
Our complete analysis for MPF will give improved or accurate cost precisely reflecting the commutator scaling for the family of quantum algorithms using interpolation or extrapolation of Trotterization.

Another important target is the MPF extended to time-dependent systems \cite{watkins-2024-clock,mizuta2024arxiv-time-dep-PF,cao2024arxiv-time-dep-PF}.
In particular, Ref. \cite{mizuta2024arxiv-time-dep-PF} has recently derived the commutator-scaling error in the time-dependent MPF, but it relies on the convergence of the BCH formula as well.
In contrast to time-independent cases, the error analysis involves product formulas composed of unbounded Hamiltonians.
However, the asymptotic convergence of the BCH formula is relevant to the nested commutators rather than the Hamiltonian itself.
We expect that our technique also applies to the time-dependent MPF and will correctly provide the near-optimal cost brought by the commutator scaling.

\subsection{Discussion and Summary}

In this paper, we discuss the problem of the commutator-scaling error of the MPF.
Although it seems to have been resolved by the previous result by Theorem \ref{Thm:original} \cite{aftab2024-mpf}, we point out that it is incompatible with the cost reflecting the commutator scaling.
We derive the alternative error bound of MPF, Theorem \ref{Thm:ours_informal}, which has a much looser requirement and a better error bound having the truncation order.
Consequently, we prove the query complexity proportional to $\tilde{O}(N^{\frac1{p+1}})$, matching with the commutator scaling, keeping the polylogarithmic dependence in $1/\varepsilon$.
We expect that our error bound does not seem to be still an optimal one for the generic order MPF; although it is natural to expect that the order-$m$ MPF $M_{pmJ}(\tau)=e^{-iH\tau}+\order{\tau^{m+1}}$ involves at most $(m+1)$-fold nested commutators in its error, our truncation order $p_0(N,\epsilon) \in \Theta (\log (N/\epsilon))$ can be independently large.
On the other hand, when focusing on Hamiltonian simulation with the well-conditioned MPF, the order $m \in \Theta (\log (Ngt/\varepsilon))$ [See Eq. (\ref{Eq_Pf:r_proof6})] matches the truncation order only by a constant factor. 
Our error bound seems to be an optimal one for Hamiltonian simulation.
In any case, we conclude with our result that the MPF has good computational cost both in the system size and accuracy, inheriting the advantages respectively in Trotterization and LCU.

\section*{Acknowledgment}

We thank D. An, who is one of the authors of Ref. \cite{aftab2024-mpf}, for sincerely responding to our question about the error bound of MPF. 
We also thank T. N. Ikeda, K. Fujii, and T. Mori for fruitful discussion.
K. M. is supported by JST PRESTO Grant No. JPMJPR235A, JSPS KAKENHI Grant No. JP24K16974, and  JST Moonshot R\&D Grant No. JPMJMS2061.

\bibliographystyle{quantum}
\bibliography{bibliography.bib}

\begin{thebibliography}{10}

\bibitem{Suzuki1991-za}
Masuo Suzuki.
\newblock ``General theory of fractal path integrals with applications to
  many‐body theories and statistical physics''.
\newblock \href{https://dx.doi.org/https://doi.org/10.1063/1.529425}{J. Math.
  Phys. {\bf 32}, 400--407}~(1991).

\bibitem{Lloyd1996-ko}
Seth Lloyd.
\newblock ``{Universal Quantum Simulators}''.
\newblock
  \href{https://dx.doi.org/https://doi.org/10.1126/science.273.5278.1073}{Science
  {\bf 273}, 1073--1078}~(1996).

\bibitem{Somma2015-fs-commutator}
R~Somma.
\newblock ``A trotter-suzuki approximation for lie groups with applications to
  hamiltonian simulation''.
\newblock \href{https://dx.doi.org/https://doi.org/10.1063/1.4952761}{Journal
  of Mathematical Physics {\bf 57}, 062202}~(2015).

\bibitem{childs-prl2019-pf}
Andrew~M. Childs and Yuan Su.
\newblock ``Nearly optimal lattice simulation by product formulas''.
\newblock \href{https://dx.doi.org/10.1103/PhysRevLett.123.050503}{Phys. Rev.
  Lett. {\bf 123}, 050503}~(2019).

\bibitem{childs2021-trotter}
Andrew~M. Childs, Yuan Su, Minh~C. Tran, Nathan Wiebe, and Shuchen Zhu.
\newblock ``Theory of trotter error with commutator scaling''.
\newblock \href{https://dx.doi.org/10.1103/PhysRevX.11.011020}{Phys. Rev. X
  {\bf 11}, 011020}~(2021).

\bibitem{Berry-prl2015-LCU}
Dominic~W. Berry, Andrew~M. Childs, Richard Cleve, Robin Kothari, and
  Rolando~D. Somma.
\newblock ``{Simulating Hamiltonian Dynamics with a Truncated Taylor Series}''.
\newblock \href{https://dx.doi.org/10.1103/PhysRevLett.114.090502}{Phys. Rev.
  Lett. {\bf 114}, 090502}~(2015).

\bibitem{Low2017-qsp}
Guang~Hao Low and Isaac~L Chuang.
\newblock ``{Optimal Hamiltonian Simulation by Quantum Signal Processing}''.
\newblock
  \href{https://dx.doi.org/https://doi.org/10.1103/PhysRevLett.118.010501}{Phys.
  Rev. Lett. {\bf 118}, 010501}~(2017).

\bibitem{Low2019-qubitization}
Guang~Hao Low and Isaac~L Chuang.
\newblock ``Hamiltonian simulation by qubitization''.
\newblock
  \href{https://dx.doi.org/https://doi.org/10.22331/q-2019-07-12-163}{Quantum
  {\bf 3}, 163}~(2019).

\bibitem{Gilyen2019-qsvt}
Andr{\'a}s Gily{\'e}n, Yuan Su, Guang~Hao Low, and Nathan Wiebe.
\newblock ``Quantum singular value transformation and beyond: exponential
  improvements for quantum matrix arithmetics''.
\newblock In Proceedings of the 51st Annual {ACM} {SIGACT} Symposium on Theory
  of Computing.
\newblock
  \href{https://dx.doi.org/https://doi.org/10.1145/3313276.3316366}{Pages
  193--204}.
\newblock STOC 2019~(2019).

\bibitem{Childs_Wiebe_2012_mpf}
Andrew~M. Childs and Nathan Wiebe.
\newblock ``Hamiltonian simulation using linear combinations of unitary
  operations''.
\newblock \href{https://dx.doi.org/10.26421/qic12.11-12}{Quantum Info. Comput.
  {\bf 12}, 901–924}~(2012).

\bibitem{low2019-mpf}
Guang~Hao Low, Vadym Kliuchnikov, and Nathan Wiebe.
\newblock ``{Well-conditioned multiproduct Hamiltonian simulation}''~(2019).
\newblock  \href{http://arxiv.org/abs/1907.11679}{arXiv:1907.11679}.

\bibitem{Faehrmann_2022-mpf}
Paul~K. Faehrmann, Mark Steudtner, Richard Kueng, Maria Kieferova, and Jens
  Eisert.
\newblock ``{Randomizing multi-product formulas for Hamiltonian simulation}''.
\newblock \href{https://dx.doi.org/10.22331/q-2022-09-19-806}{Quantum {\bf 6},
  806}~(2022).

\bibitem{Vazquez2022-mpf}
Almudena~Carrera Vazquez, Ralf Hiptmair, and Stefan Woerner.
\newblock ``{Enhancing the quantum linear systems algorithm using Richardson
  extrapolation}''.
\newblock \href{https://dx.doi.org/10.1145/3490631}{ACM Transactions on Quantum
  Computing {\bf 3}, 1--37}~(2022).

\bibitem{Carrera_Vazquez_2023-mpf}
Almudena Carrera~Vazquez, Daniel~J. Egger, David Ochsner, and Stefan Woerner.
\newblock ``Well-conditioned multi-product formulas for hardware-friendly
  hamiltonian simulation''.
\newblock \href{https://dx.doi.org/10.22331/q-2023-07-25-1067}{Quantum {\bf 7},
  1067}~(2023).

\bibitem{zhuk2024-mpf}
Sergiy Zhuk, Niall~F. Robertson, and Sergey Bravyi.
\newblock ``{Trotter error bounds and dynamic multi-product formulas for
  Hamiltonian simulation}''.
\newblock \href{https://dx.doi.org/10.1103/PhysRevResearch.6.033309}{Phys. Rev.
  Res. {\bf 6}, 033309}~(2024).

\bibitem{aftab2024-mpf}
Junaid Aftab, Dong An, and Konstantina Trivisa.
\newblock ``{Multi-product Hamiltonian simulation with explicit commutator
  scaling}''~(2024).
\newblock  \href{http://arxiv.org/abs/2403.08922}{arXiv:2403.08922}.

\bibitem{Abanin2015-zg}
Dmitry~A Abanin, Wojciech De~Roeck, and Fran{\c c}ois Huveneers.
\newblock ``{Exponentially Slow Heating in Periodically Driven Many-Body
  Systems}''.
\newblock
  \href{https://dx.doi.org/https://doi.org/10.1103/PhysRevLett.115.256803}{Phys.
  Rev. Lett. {\bf 115}, 256803}~(2015).

\bibitem{Kuwahara2016-yn}
Tomotaka Kuwahara, Takashi Mori, and Keiji Saito.
\newblock ``{Floquet–Magnus} theory and generic transient dynamics in
  periodically driven many-body quantum systems''.
\newblock
  \href{https://dx.doi.org/https://doi.org/10.1016/j.aop.2016.01.012}{Ann.
  Phys. {\bf 367}, 96--124}~(2016).

\bibitem{Mori2016-tp}
Takashi Mori, Tomotaka Kuwahara, and Keiji Saito.
\newblock ``{Rigorous Bound on Energy Absorption and Generic Relaxation in
  Periodically Driven Quantum Systems}''.
\newblock
  \href{https://dx.doi.org/https://doi.org/10.1103/PhysRevLett.116.120401}{Phys.
  Rev. Lett. {\bf 116}, 120401}~(2016).

\bibitem{Abanin2017-li}
Dmitry~A Abanin, Wojciech De~Roeck, Wen~Wei Ho, and Fran{\c c}ois Huveneers.
\newblock ``{Effective Hamiltonians, prethermalization, and slow energy
  absorption in periodically driven many-body systems}''.
\newblock
  \href{https://dx.doi.org/https://doi.org/10.1103/PhysRevB.95.014112}{Phys.
  Rev. B {\bf 95}, 014112}~(2017).

\bibitem{Abanin2017-zs}
Dmitry Abanin, Wojciech De~Roeck, Wen~Wei Ho, and Fran{\c c}ois Huveneers.
\newblock ``A rigorous theory of {Many-Body} prethermalization for periodically
  driven and closed quantum systems''.
\newblock
  \href{https://dx.doi.org/https://doi.org/10.1007/s00220-017-2930-x}{Commun.
  Math. Phys. {\bf 354}, 809--827}~(2017).

\bibitem{Endo-PRA2019-mitigation}
Suguru Endo, Qi~Zhao, Ying Li, Simon Benjamin, and Xiao Yuan.
\newblock ``{Mitigating algorithmic errors in a Hamiltonian simulation}''.
\newblock \href{https://dx.doi.org/10.1103/PhysRevA.99.012334}{Phys. Rev. A
  {\bf 99}, 012334}~(2019).

\bibitem{Rendon2023-interpolate}
Gumaro Rendon.
\newblock ``{All you need is Trotter}''~(2023).
\newblock  \href{http://arxiv.org/abs/2311.01533}{arXiv:2311.01533}.

\bibitem{Rendon2024-interpolate}
Gumaro Rendon, Jacob Watkins, and Nathan Wiebe.
\newblock ``{Improved accuracy for Trotter simulations using Chebyshev
  interpolation}''.
\newblock
  \href{https://dx.doi.org/https://doi.org/10.22331/q-2024-02-26-1266}{Quantum
  {\bf 8}, 1266}~(2024).

\bibitem{Watson2024-mpf}
James~D. Watson and Jacob Watkins.
\newblock ``{Exponentially Reduced Circuit Depths Using Trotter Error
  Mitigation}''.
\newblock \href{https://dx.doi.org/10.1103/kw39-yxq5}{PRX Quantum {\bf 6},
  030325}~(2025).

\bibitem{watson2024-mpf-random}
James~D. Watson.
\newblock ``{Randomly Compiled Quantum Simulation with Exponentially Reduced
  Circuit Depths}''~(2024).
\newblock  \href{http://arxiv.org/abs/2411.04240}{arXiv:2411.04240}.

\bibitem{chakraborty2025-mpf}
Shantanav Chakraborty, Soumyabrata Hazra, Tongyang Li, Changpeng Shao, Xinzhao
  Wang, and Yuxin Zhang.
\newblock ``Quantum singular value transformation without block encodings:
  Near-optimal complexity with minimal ancilla''~(2025).
\newblock  \href{http://arxiv.org/abs/2504.02385}{arXiv:2504.02385}.

\bibitem{Guo2025-interpolate}
Taozhi Guo, Gumaro Rendon, and Rutuja Kshirsagar.
\newblock ``{Canonical Partition Function on a Quantum Computer through Trotter
  Interpolation}''~(2025).
\newblock  \href{http://arxiv.org/abs/2506.09318}{arXiv:2506.09318}.

\bibitem{mizuta2025_low_energy}
Kaoru Mizuta and Tomotaka Kuwahara.
\newblock ``{Trotterization is Substantially Efficient for Low-Energy
  States}''.
\newblock \href{https://dx.doi.org/10.1103/q87n-5xhz}{Phys. Rev. Lett. {\bf
  135}, 130602}~(2025).

\bibitem{Chin2010-mpf}
Siu~A Chin.
\newblock ``Multi-product splitting and {Runge-Kutta-Nystr{\"o}m}
  integrators''.
\newblock
  \href{https://dx.doi.org/https://doi.org/10.1007/s10569-010-9255-9}{Celest.
  Mech. Dyn. Astron. {\bf 106}, 391--406}~(2010).

\bibitem{Brassard_2002_qaa}
Gilles Brassard, Peter H{\o}yer, Michele Mosca, and Alain Tapp.
\newblock ``Quantum amplitude amplification and estimation''.
\newblock
  \href{https://dx.doi.org/https://doi.org/10.1090/conm/305/05215}{{Quantum
  Computation and Information}Page 53–74}~(2002).

\bibitem{Blanes2009-op-magnus}
S~Blanes, F~Casas, J~A Oteo, and J~Ros.
\newblock ``{The Magnus expansion and some of its applications}''.
\newblock \href{https://dx.doi.org/10.1016/j.physrep.2008.11.001}{Phys. Rep.
  {\bf 470}, 151--238}~(2009).

\bibitem{Lazarides2014-iv-eth}
Achilleas Lazarides, Arnab Das, and Roderich Moessner.
\newblock ``Equilibrium states of generic quantum systems subject to periodic
  driving''.
\newblock \href{https://dx.doi.org/10.1103/PhysRevE.90.012110}{Phys. Rev. E
  {\bf 90}, 012110}~(2014).

\bibitem{DAlessio2014-iu-eth}
Luca D’Alessio and Marcos Rigol.
\newblock ``Long-time behavior of isolated periodically driven interacting
  lattice systems''.
\newblock \href{https://dx.doi.org/10.1103/PhysRevX.4.041048}{Phys. Rev. X {\bf
  4}, 041048}~(2014).

\bibitem{Sharma2024-om-magnus}
Kunal Sharma and Minh~C Tran.
\newblock ``{Hamiltonian Simulation in the Interaction Picture Using the Magnus
  Expansion}''~(2024).
\newblock  \href{http://arxiv.org/abs/2404.02966}{arXiv:2404.02966}.

\bibitem{arnal2020-bch}
Ana Arnal, Fernando Casas, and Cristina Chiralt.
\newblock ``{A Note on the Baker--Campbell--Hausdorff Series in Terms of
  Right-Nested Commutators}''.
\newblock \href{https://dx.doi.org/10.1007/s00009-020-01681-6}{Mediterranean
  Journal of Mathematics {\bf 18}, 53}~(2021).

\bibitem{Haah2021-time-dep}
Jeongwan Haah, Matthew~B Hastings, Robin Kothari, and Guang~Hao Low.
\newblock ``{Quantum algorithm for simulating real time evolution of lattice
  Hamiltonians}''.
\newblock \href{https://dx.doi.org/https://doi.org/10.1137/18M1231511}{SIAM J.
  Comput.Pages FOCS18--250--FOCS18--284}~(2021).

\bibitem{Lieb1972-uo}
Elliott~H Lieb and Derek~W Robinson.
\newblock ``The finite group velocity of quantum spin systems''.
\newblock \href{https://dx.doi.org/https://doi.org/10.1007/BF01645779}{Commun.
  Math. Phys. {\bf 28}, 251--257}~(1972).

\bibitem{Tran-PRX2019-hhkl}
Minh~C. Tran, Andrew~Y. Guo, Yuan Su, James~R. Garrison, Zachary Eldredge,
  Michael Foss-Feig, Andrew~M. Childs, and Alexey~V. Gorshkov.
\newblock ``Locality and digital quantum simulation of power-law
  interactions''.
\newblock \href{https://dx.doi.org/10.1103/PhysRevX.9.031006}{Phys. Rev. X {\bf
  9}, 031006}~(2019).

\bibitem{watkins-2024-clock}
Jacob Watkins, Nathan Wiebe, Alessandro Roggero, and Dean Lee.
\newblock ``{Time-Dependent Hamiltonian Simulation Using Discrete-Clock
  Constructions}''.
\newblock \href{https://dx.doi.org/10.1103/PRXQuantum.5.040316}{PRX Quantum
  {\bf 5}, 040316}~(2024).

\bibitem{mizuta2024arxiv-time-dep-PF}
Kaoru Mizuta, Tatsuhiko~N. Ikeda, and Keisuke Fujii.
\newblock ``Explicit error bounds with commutator scaling for time-dependent
  product and multi-product formulas''~(2024).
\newblock  \href{http://arxiv.org/abs/2410.14243}{arXiv:2410.14243}.

\bibitem{cao2024arxiv-time-dep-PF}
Yu~Cao, Shi Jin, and Nana Liu.
\newblock ``Unifying framework for quantum simulation algorithms for
  time-dependent hamiltonian dynamics''.
\newblock \href{https://dx.doi.org/10.1103/fkh5-b669}{Phys. Rev. Res. {\bf 7},
  043186}~(2025).

\end{thebibliography}

\onecolumn\newpage
\appendix

\section{Bounds on the nested commutators and the BCH expansions}\label{SubsecA:bounds}

In this appendix, we prove some inequalities on the nested commutators and the operator $\Phi_q$ in the BCH expansion, Eq. (\ref{Eq_Pf:BCH}), which is composed of the nested commutators.
First, we prove the following relation about the nested commutators.

\begin{lemma}\label{LemmaA:nested_bound}
\textbf{(Nested commutators)}

Let $H=\sum_\gamma H_\gamma$ be a $k$-local and $g$-extensive Hamiltonian.
When an operator $O_X$ nontrivially acts on a domain $X$ with $|X| \leq k$, the following inequality on the nested commutator holds, 
\begin{equation}\label{EqA_A0:nested_bound}
    \sum_{\gamma_1,\cdots,\gamma_q=1}^\Gamma \norm{[H_{\gamma_q},\cdots,[H_{\gamma_{q'+1}},[O_X,[H_{\gamma_{q'}},\cdots,[H_{\gamma_2},H_{\gamma_1}]]]]]} \leq q! (2kg)^q \norm{O_X}. 
\end{equation}
for any $q'=1,\cdots,q$.
\end{lemma}

\textbf{Proof.---} The difference from the inequality, Eq. (\ref{Eq_Pre:commutator_bound}), is the position of the local term $h_i^\gamma$, which is confined to a domain around the site $i$.
The proof is similar to the one for Theorem S1 in Ref. \cite{Kuwahara2016-yn}.
We expand the nested commutator in Eq. (\ref{EqA_A0:nested_bound}) based on Eq. (\ref{Eq_Pre:H_partition}), which results in
\begin{eqnarray}
    && [H_{\gamma_q},\cdots,[H_{\gamma_{q'+1}},[O_X,[H_{\gamma_{q'}},\cdots,[H_{\gamma_2},H_{\gamma_1}]]]]] \nonumber \\
    && \qquad = \sum_{\gamma_1,\cdots,\gamma_q}\sum_{X_1,\cdots,X_q \subset \Lambda} [h_{X_q}^{\gamma_q},\cdots,[h_{X_{q'+1}}^{\gamma_{q'+1}},[O_X,[h_{X_{q'}}^{\gamma_{q'}},\cdots,[h_{X_2}^{\gamma_2},h_{X_1}^{\gamma_1}]]]]].
\end{eqnarray}
In the right-hand side, nested commutators among local terms having nontrivial overlaps can survive.
Let us define the union $Y_{q''} = X_1 \cup \cdots \cup X_{q''}$ for $q''=1,2,\cdots,q$.
Nested commutators among the local terms survive only for the set of domains $(X_1,\cdots,X_q)$ satisfying
\begin{equation}
    \begin{cases}
        X_{q''} \cap Y_{q''-1} \neq \phi & (q''=2,\cdots,q') \\
        Y_{q'} \cap X \neq \phi \\
        X_{q''} \cap (Y_{q''-1} \cup X) \neq \phi & (q''=q'+1,\cdots,q)
    \end{cases},
\end{equation}
which we denote as $\mcl{X}_{q'}(X)$.
The left-hand side of Eq. (\ref{EqA_A0:nested_bound}) is expressed by
\begin{eqnarray}
    && \sum_{\gamma_1,\cdots,\gamma_q} \norm{[H_{\gamma_q},\cdots,[H_{\gamma_{q'+1}},[O_X,[H_{\gamma_{q'}},\cdots,[H_{\gamma_2},H_{\gamma_1}]]]]]} \nonumber \\
    && \quad = \sum_{\gamma_1,\cdots,\gamma_q} \norm{ \sum_{\substack{X_1,\cdots,X_q \subset \Lambda: \\ (X_1,\cdots,X_q) \in \mcl{X}_{q'}(X)}}  [h_{X_q}^{\gamma_q},\cdots,[h_{X_{q'+1}}^{\gamma_{q'+1}},[O_X,[h_{X_{q'}}^{\gamma_{q'}},\cdots,[h_{X_2}^{\gamma_2},h_{X_1}^{\gamma_1}]]]]]} \nonumber \\
    && \quad \leq 2^q \norm{O_X} \sum_{\substack{X_1,\cdots,X_q \subset \Lambda: \\ (X_1,\cdots,X_q) \in \mcl{X}_{q'}(X)}}  \prod_{q''=1}^q \left( \sum_{\gamma_{q''}}\norm{h_{X_{q''}}^{\gamma_{q''}}}\right). \label{EqA_A0:proof1}
\end{eqnarray}

We first consider the summation over $X_{q''}$ for $q''=q'+1,\cdots,q$, in which each $X_{q''}$ has a nonempty intersection with $Y_{q''-1} \cup X$.
Since each domain $Y_{q''-1} \cup X$ contains at most $q''k$ sites due to the $k$-locality, we have
\begin{eqnarray}
    \sum_{\substack{X_{q''} \subset \Lambda: \\ X_{q''} \cap (Y_{q''-1} \cap X) \neq \phi}} \sum_{\gamma_{q''}} \norm{h_{X_{q''}}^{\gamma_{q''}}} &\leq& \sum_{j \in Y_{q''-1} \cap X} \sum_{X_{q''} \ni j} \sum_{\gamma_{q''}}\norm{h_{X_{q''}}^{\gamma_{q''}}} \nonumber \\
    &\leq& q'' kg \label{EqA_A0:proof2}
\end{eqnarray}
for each $q''=q'+1,\cdots,q$.
As a result, Eq. (\ref{EqA_A0:proof1}) is further bounded by
\begin{eqnarray}
    [\text{Eq. (\ref{EqA_A0:proof1})}] &\leq& 2^q \norm{O_X} \sum_{\substack{X_1,\cdots,X_{q'} \subset \Lambda: \\ X_{q''} \cap Y_{q''-1} \neq \phi \, (q''=2,\cdots,q'), \\
    Y_{q'} \cap X \neq \phi}} \prod_{q''=1}^{q'} \left(\sum_{\gamma_{q''}}\norm{h_{X_{q''}}^{\gamma_{q''}}}\right) \frac{q!}{q'!} (kg)^{q-q'}. 
\end{eqnarray}
In the above formula, the summation over $X_1,\cdots,X_{q'}$ has an upper bound which can be evaluated in a recursive way.
Equations (S17) and (S19) in Ref. \cite{mizuta2025_low_energy} give the relation,
\begin{equation}
    \sum_{\substack{X_1,\cdots,X_{q'} \subset \Lambda: \\ X_{q''} \cap Y_{q''-1} \neq \phi \, (q''=2,\cdots,q'), \\
    Y_{q'} \cap X \neq \phi}} \prod_{q''=1}^{q'} \left(\sum_{\gamma_{q''}}\norm{h_{X_{q''}}^{\gamma_{q''}}}\right) \leq q' ! (kg)^{q'}.
\end{equation}
Thus, we can conclude Eq. (\ref{EqA_A0:nested_bound}). $\quad \square$

This lemma reproduces Lemma 3 in Ref. \cite{Kuwahara2016-yn} with $q'=0$.
We use it for deriving the extensiveness of the BCH expansion $\Phi_q$ (below) and evaluating the series expansions of the subsystem operators (Appendix \ref{SubsecA:Series_expansion}).

\begin{lemma}\label{LemmaA:locality_extensive_BCH}
\textbf{(Locality and extensiveness of the BCH expansion)}

Let $H=\sum_\gamma H_\gamma$ be a $k$-local and $g$-extensive Hamiltonian.
Then, the operator $\Phi_q$, i.e., the order-$q$ term of the BCH expansion defined by Eq. (\ref{Eq_Pf:BCH}), has the following locality and extensiveness,
\begin{equation}
    \text{locality:} \quad qk, \qquad \text{extensiveness:} \quad \frac{(q-1)!}{q} (2c_pkg)^{q-1} c_pg.
\end{equation}
\end{lemma}

\textbf{Proof.---}
The locality $qk$ is clear from the fact that $\Phi_q$ is composed of $q$-tuple nested commutators among $k$-local operators.
We concentrate on the extensiveness of $\Phi_q$, denoted by $g(\Phi_q)$.
Using the expansion $H=\sum_X h_X$, the operator $\Phi_q$ is expressed by the nested commuters among $h_{X_1}^1,\cdots,h_{X_{q_1}}^1,\cdots,h_{X_{q-q_{c_p\Gamma}+1}}^{c_p\Gamma}, \cdots, h_{X_q}^{c_p\Gamma}$.
Nontrivial terms nontrivially acting on an arbitrary site $j \in \Lambda$ should involve a domain $X_{q'}$ such that $X_{q'} \ni j$.
Thus, the extensiveness of $\Phi_q$ is bounded by
\begin{eqnarray}
    g(\Phi_q) &\leq& \sum_{\substack{q_1,\cdots,q_{c_p \Gamma } \geq 0 \\ q_1+\cdots+q_{c_p\Gamma} = q }} \frac1{q_1!\cdots q_{c_p\Gamma}!} \nonumber \\
    && \qquad \times \sum_{q'=1}^q \sum_{\substack{X_1,\cdots,X_q: \\ X_{q'} \ni j}} \norm{\phi(\underbrace{h_{X_1}^1,\cdots,h_{X_{q_1}}^1}_{q_1},\cdots,\underbrace{h_{X_{q-q_{c_p\Gamma}+1}}^{c_p\Gamma}, \cdots, h_{X_q}^{c_p\Gamma}}_{q_{c_p\Gamma}})} \nonumber \\
    &\leq& \frac{1}{q^2} \sum_{\substack{q_1,\cdots,q_{c_p \Gamma } \geq 0 \\ q_1+\cdots+q_{c_p\Gamma} = q }} \frac1{q_1!\cdots q_{c_p\Gamma}!} \nonumber \\
    && \quad \times \sum_{\sigma \in S_q} \frac{1}{\left( \begin{array}{c}
         q-1 \\
         d_\sigma
    \end{array}\right)}  \max_{\sigma \in S_q} \sum_{q'=1}^q \sum_{\substack{X_1,\cdots,X_q: \\ X_{q'} \ni j}} \norm{[\tilde{h}_{X_{\sigma(1)}}^{\sigma(1)}, \cdots, [\tilde{h}_{X_{\sigma(q-1)}}^{\sigma(q-1)},\tilde{h}_{X_{\sigma(q)}}^{\sigma(q)}]]}.
\end{eqnarray}
In the above, each term $\tilde{h}_{X_{q'}}^{q'}$ means one of the local terms $\{h_X\}$, which is defined so that the indices in the subscript and superscript match in the first line as follows,
\begin{equation}
    \tilde{h}_{X_{q'}}^{q'} = \begin{cases}
        h_{X_{q'}}^1 & 1 \leq q' \leq q_1 \\
        h_{X_{q'}}^2 & q_1 < q' \leq q_1+q_2 \\
        \vdots \\
        h_{X_{q'}}^{c_p \Gamma} & q-q_{c_p\Gamma} < q' \leq q.
    \end{cases}
\end{equation}
We then use the inequality, $\sum_{\sigma \in S_q} \left( \begin{array}{c}
     q-1 \\
     d_\sigma
\end{array}\right)^{-1} \leq q_1!\cdots q_{c_p\Gamma}!$ under $q_1+\cdots+q_{c_p\Gamma}=q$.
We also note that each $\tilde{h}_{X_{\sigma(q')}}^{\sigma(q')}$ reproduces $c_p$ copies of $h_{X_{\sigma(q')}}^\gamma$ for each $\gamma$ in the summation over $q_1,\cdots,q_{c_p\Gamma}$ with $q_1+\cdots+q_{c_p\Gamma}=q$, just like the commutator-scaling error in Trotterization \cite{childs2021-trotter}.
Thus, the extensiveness of $\Phi_q$ is further bounded by
\begin{eqnarray}
    g(\Phi_q) &\leq& \frac{(c_p)^q}{q^2} \sum_{\gamma_1,\cdots,\gamma_q} \sum_{q'=1}^q \sum_{\substack{X_1,\cdots,X_q: \\ X_{q'} \ni j}} \max_{\sigma \in S_q} \norm{[h_{X_{\sigma(1)}}^{\gamma_{\sigma(1)}}, \cdots, [h_{X_{\sigma(q-1)}}^{\gamma_{\sigma(q-1)}},h_{X_{\sigma(q)}}^{\gamma_{\sigma(q)}}]]} \nonumber \\
    &\leq& \frac{(c_p)^q}{q^2} \sum_{q'=1}^q \sum_{\gamma_{q'}} \sum_{X_{q'} \ni j} (q-1)! (2kg)^{q-1} \norm{h_{X_{q'}}^{\gamma_{q'}}} \nonumber \\
    &\leq& \frac{(q-1)!}q (2c_pkg)^{q-1} c_pg.
\end{eqnarray}
The second inequality comes from the fact that the nested commutators in the first line share the same upper bound by Lemma \ref{LemmaA:nested_bound} regardless of the permutation $\sigma \in S_q$, where we fix the domain $X_{q'}$.
$\quad \square$

\section{Series expansions of the subsystem Trotterization and BCH formula}\label{SubsecA:Series_expansion}

Here, we prove Eq. (\ref{Eq_Pf:series_expansion_result}) in the main text, which is used for the derivation of Theorem \ref{Thm:error_truncated_BCH}.
They provide the upper bounds on the series expansions for the subsystem operators, and essentially arise from the locality and the extensiveness of Hamiltonians.

\subsection{Series expansion of the subsystem Trotterization}

We begin with proving the following bound about the subsystem Trotterization $T^{>i}_p(\tau)$ defined by Eq. (\ref{Eq_Pf:subsys_Trot_def}).

\begin{lemma}\label{LemmaA:Series_subsys_Trot}
\textbf{}

Let us expand the operator $T^{>i}_p(\tau)^\dagger T^{>i-1}_p(\tau)$ by
\begin{equation}\label{EqA:T_i_expansion}
    T^{>i}_p(\tau)^\dagger T^{>i-1}_p(\tau) = \sum_{q=0}^\infty T_q^i \tau^q,
\end{equation}
where the subsystem Trotterization $T^{>i}_p(\tau)$ is defined by Eq. (\ref{Eq_Pf:subsys_Trot_def}).
Then, we have
\begin{equation}\label{EqA:T_n_bound}
    \norm{T_q^i} \leq (4c_p kg)^q.
\end{equation}
\end{lemma}

\textbf{Proof.---}
The operator $T_p^{>i}(\tau)$ defined by Eq. (\ref{Eq_Pf:subsys_Trot_def}) can be seen as a time evolution operator under a Hamiltonian,
\begin{equation}\label{EqA:Trotter_periodic}
H^{>i} (\tau') = \begin{cases}
        \alpha_1 H_{\gamma_1}^{>i} & \text{if $\tau' \in [0,\tau)$} \\
        \vdots \\
        \alpha_v H_{\gamma_v}^{>i} & \text{if $\tau' \in [(v-1)\tau,v\tau)$} \\
        \vdots \\
        \alpha_{c_p\Gamma} H_{\gamma_{c_p\Gamma}}^{>i} & \text{if $\tau' \in [(c_p\Gamma-1)\tau,c_p\Gamma\tau)$}
    \end{cases},
\end{equation}
over the time $\tau' \in [0,c_p\Gamma \tau]$.
We apply the interaction picture under $H^{>i} (\tau')$ to the Hamiltonian $H^{>i-1} (\tau') = h_i(\tau') + H^{>i} (\tau')$, where we define $h_i(\tau')=H^{>i-1} (\tau') - H^{>i} (\tau')$.
We note that $h_i(\tau')$ is a partial set of local terms acting on the site $i$, whose expression at instantaneous time is given by
\begin{equation}\label{EqA:h_i_tau_def}
    h_i(\tau') =  \alpha_v \sum_{\substack{X: X \ni i, \\ X \cap \{1,\cdots,i-1\} = \phi}} h_X^{\gamma_v}, \quad \text{for $\tau' \in [(v-1)\tau,v\tau)$}
\end{equation}
based on Eqs. (\ref{Eq_Pf:subsys_H_def}) and (\ref{EqA:Trotter_periodic}).
The operator $T_p^{>i-1}(\tau)$ is expressed by
\begin{eqnarray}
    T_p^{>i-1}(\tau) &=& T_p^{>i}(\tau) \times \mcl{T} \exp \left( - i \int_0^{c_p\Gamma \tau} \dd \tau' U^{>i}(\tau')^\dagger h_i (\tau') U^{>i}(\tau') \right), \label{EqA:TT_UhU} \\
    U^{>i}(\tau') &=& \mcl{T} \exp \left( - i \int_0^{\tau'} \dd \tau'' H^{>i}(\tau'') \right).
\end{eqnarray}
We apply the Dyson series expansion on the operator $U^{>i}(\tau')^\dagger h_i (\tau') U^{>i}(\tau')$, which results in 
\begin{eqnarray}
    && U^{>i}(\tau')^\dagger h_i (\tau') U^{>i}(\tau') \nonumber \\
    && \qquad = \sum_{q=0}^\infty \int_0^{\tau'} \dd \tau_1' \cdots \int_0^{\tau_{q-1}'} \dd \tau_q' \left\{\prod_{q'=1,\cdots,q}^\leftarrow \left( i \ad_{H^{>i}(\tau_{q'})} \right)\right\} h_i(\tau') \nonumber \\
    && \qquad =  \sum_{q=0}^\infty (\tau')^q \times \underbrace{\frac{i^q}{q!} \int_0^1 \dd s_1 \cdots \int_0^1 \dd s_q \mcl{T}_\leftarrow \left[ \prod_{q'=1}^q \left( \ad_{H^{>i}(s_{q'}\tau')}\right) \right] h_i(\tau')}_{\Xi_q (\tau')}. \label{EqA:Xi_n_def}
\end{eqnarray}
The symbol $\mcl{T}_\leftarrow [\cdot]$ denotes the time-ordered product, in which the variables $s_1,\cdots,s_q$ are aligned in the descending order.
We again apply the Dyson series expansion to the operator $T^{>i}_p(\tau)^\dagger T^{>i-1}_p(\tau)$.
Equation (\ref{EqA:TT_UhU}) immediately leads to the expression,
\begin{eqnarray}
    && T^{>i}_p(\tau)^\dagger T^{>i-1}_p(\tau) \nonumber \\
    && \quad = \sum_{n=0}^\infty (-i)^n \int_0^{c_p \Gamma\tau} \dd \tau_1 \int_0^{\tau_1} \dd \tau_2 \cdots \int_0^{\tau_{n-1}} \dd \tau_n \prod_{n'=1,\cdots,n}^\rightarrow \left( \sum_{q_{n'}=0}^\infty (i\tau_{n'})^{q_{n'}} \Xi_{q_{n'}}(\tau_{n'})\right) \nonumber \\
    && \quad = \sum_{n=0}^\infty (-i)^n \sum_{q_1,\cdots,q_n=0}^\infty \frac{\tau^{n+q_1+\cdots+q_n}}{n!} \int_0^{c_p\Gamma} \dd s_1 \cdots \int_0^{c_p\Gamma} \dd s_n \mcl{T}_\rightarrow \left[ \prod_{n'=1}^n (s_{n'})^{q_{n'}} \Xi_{q_{n'}}(s_{n'}\tau)\right] \nonumber \\
    && \quad = \sum_{q=0}^\infty \tau^q \underbrace{\sum_{n=0}^q \frac1{n!} \sum_{\substack{q_1,\cdots,q_n \geq 0: \\ q_1+\cdots+q_n=q-n}} \int_0^{c_p\Gamma} \dd s_1 \cdots \int_0^{c_p\Gamma} \dd s_n \mcl{T}_\rightarrow \left[ \prod_{n'=1}^n (s_{n'})^{q_{n'}} \Xi_{q_{n'}}(s_{n'}\tau)\right]}_{T_q^i}. \label{EqA:T_n_def}
\end{eqnarray}
The symbol $\mcl{T}_\rightarrow [\cdot]$ is defined in the opposite way of $\mcl{T}_\leftarrow [\cdot]$.
This gives the series expansion of the operator $T^{>i}_p(\tau)^\dagger T^{>i-1}_p(\tau)$ in Eq. (\ref{EqA:T_i_expansion}).
We can confirm that the coefficient $T_q^i$ defined by Eq. (\ref{EqA:T_n_def}) is independent of the time $\tau$ in the following way.
The operator $T_q^i$ appears to have the time-dependency in $H^{>i}(s_{n'}s_{q'}\tau)$ and $h_i(s_{n'}\tau)$ via $\Xi_{q_{n'}}(s_{n'}\tau)$ [See Eq. (\ref{EqA:Xi_n_def})].
However, since they are piecewise constant modulo $\tau$ as shown in Eqs. (\ref{EqA:Trotter_periodic}) and (\ref{EqA:h_i_tau_def}), they can be replaced by the functions solely dependent on $s_{n'}s_{q'}$ and $s_{n'}$ respectively.
This results in the $\tau$-independence of $T_q^i$, which ensures that Eq. (\ref{EqA:T_n_def}) gives the series expansion of $T^{>i}_p(\tau)^\dagger T^{>i-1}_p(\tau)$.

We finally evaluate the upper bound on the coefficient $T_q^i$.
We begin with the upper bound on $\Xi_q(s\tau)$ defined by Eq. (\ref{EqA:Xi_n_def}).
For $s \in (0,c_p\Gamma)$, the Hamiltonian $H^{>i}(s_{q'}s\tau)$ is piecewise constant within
\begin{equation}
    \frac{v_{q'}-1}{s} \leq s_{q'} < \frac{v_{q'}}s, \quad v_{q'} \in \bbN.
\end{equation}
Each index $v_{q'}$ takes $1,2,\cdots, \lceil s \rceil$ for $s_{q'} \in (0,1)$.
This property allows us to replace the integral in Eq. (\ref{EqA:Xi_n_def}) by the summation as follows,
\begin{eqnarray}
    \norm{\Xi_q (s\tau)} &\leq& \frac{1}{q!} \int_0^1 \dd s_1 \cdots \int_0^1 \dd s_q \norm{\mcl{T}_\rightarrow \left[ \prod_{q'=1}^q \left( \ad_{H^{>i}(s_{q'}s \tau)}\right) \right] h_i(s\tau)} \nonumber \\
    &\leq& \frac{1}{q!} \sum_{v_1,v_2,\cdots,v_q=1}^{\lceil s \rceil}\int_{\frac{v_1-1}s}^{\frac{v_1}s} \dd s_1 \cdots \int_{\frac{v_q-1}s}^{\frac{v_q}s} \dd s_q \norm{\mcl{T}_\rightarrow \left[ \prod_{q'=1}^q \left( \alpha_{v_{q'}} \ad_{H^{>i}_{\gamma_{v_{q'}}}}\right) \right] h_i(s\tau)} \nonumber \\
    &\leq& \frac{1}{q!s^q} \sum_{v_1=1}^{c_p\Gamma} \cdots \sum_{v_q =1}^{c_p\Gamma} \norm{[\alpha_{v_1} H_{\gamma_{v_1}}^{>i},[\alpha_{v_2} H_{\gamma_{v_2}}^{>i},\cdots,[\alpha_{v_q} H_{\gamma_{v_q}}^{>i}, h_i(s\tau)]\cdots]]} \nonumber \\
    &\leq& \frac{(c_p)^q}{q!s^q} \sum_{\gamma_1,\cdots,\gamma_q=1}^\Gamma \norm{[ H_{\gamma_1}^{>i},[ H_{\gamma_2}^{>i},\cdots,[ H_{\gamma_q}^{>i}, h_i(s\tau)]\cdots]]} \nonumber \\
    &\leq& \left(\frac{2c_pkg}{s} \right)^q \norm{h_i(s\tau)}. \label{EqA:Xi_n_bound}
\end{eqnarray}
The time-ordered product in the second line is about the integers $v_1,\cdots,v_q$.
The fourth line comes from the fact that every index $\gamma \in \{1,\cdots,\Gamma\}$ is repeated $c_p$ times by $\gamma_v$ when $v$ runs over $v=1,\cdots,c_p\Gamma$ by the definition of Trotterization, Eq. (\ref{Eq_Pre:Trotterization}).
The last inequality comes from Lemma \ref{LemmaA:nested_bound}.
Substituting this bound into Eq. (\ref{EqA:T_n_def}), we obtain the upper bound on the coefficient $T_q^i$ by
\begin{eqnarray}
    \norm{T_q^i} &\leq& \sum_{n=0}^q \frac1{n!} \sum_{\substack{q_1,\cdots,q_n \geq 0: \\ q_1+\cdots+q_n=q-n}} \int_0^{c_p\Gamma} \dd s_1 \cdots \int_0^{c_p\Gamma} \dd s_n  \prod_{n'=1}^n \left[ (s_{n'})^{q_{n'}} \norm{\Xi_{q_{n'}}(s_{n'}\tau)} \right] \nonumber \\
    &\leq& \sum_{n=0}^q \frac1{n!} \sum_{\substack{q_1,\cdots,q_n \geq 0: \\ q_1+\cdots+q_n=q-n}} \prod_{n'=1}^n \left[ (2c_pkg)^{q_{n'}}\int_0^{c_p \Gamma} \dd s \norm{h_i(s\tau)} \right] \nonumber \\
    &=& \sum_{n=0}^q \frac{(2c_pkg)^{q-n}}{n!} (c_p g)^n \sum_{\substack{q_1,\cdots,q_n \geq 0: \\ q_1+\cdots+q_n=q-n}} 1 \nonumber \\
    &\leq& \frac12 (4c_pkg)^q \sum_{n=0}^\infty \frac{2^{-n}}{n!} = \frac{\sqrt{e}}2 (4c_pkg)^q. \label{EqA:T_n_bound_deriv}
\end{eqnarray}
The second line comes from Eq. (\ref{EqA:Xi_n_bound}).
In the third line, we use 
\begin{equation}
    \int_0^{c_p\Gamma} \dd s \norm{h_i(s\tau)} = \sum_{v=1}^{c_p\Gamma} |\alpha_v| \norm{\sum_{\substack{X: X \ni i, \\ X \cap \{1,\cdots,i-1\} = \phi}} h_X^{\gamma_v}} \leq \sum_{v=1}^{c_p \Gamma} \sum_{X \subset \Lambda: X \ni i} \norm{h_X^{\gamma_v}} \leq c_p g
\end{equation}
from the definition, Eq. (\ref{EqA:h_i_tau_def}).
In the last inequality of Eq. (\ref{EqA:T_n_bound_deriv}), we use the relation,
\begin{equation}
\sum_{\substack{q_1,\cdots,q_n \geq 0: \\ q_1+\cdots+q_n=q-n}} 1 = \frac{(q-1)!}{(n-1)!(q-n)!} \leq 2^{q-1}.
\end{equation}
Finally, using the trivial relation $\sqrt{e} < 2$, we arrive at Eq. (\ref{EqA:T_n_bound}). $\quad \square$

\subsection{Series expansion of the subsystem BCH formula}

We next focus on the second inequality in Eq. (\ref{Eq_Pf:series_expansion_result}), which is a counterpart of Lemma \ref{LemmaA:Series_subsys_Trot} for the subsystem BCH formula, Eq. (\ref{Eq_Pf:subsys_BCH_def}).
We exploit the locality and extensiveness of the effective Hamiltonian $H^{>i} + \sum_{q=2}^{p_0} \Phi_q^{>i} \tau^{q-1}$ for the BCH expansion, represented by Lemma \ref{LemmaA:locality_extensive_BCH}.
As a result, we can prove the following lemma.

\begin{lemma}\label{LemmaA:Series_subsys_BCH}
\textbf{}

Let $\tilde{T}_{p,p_0}^{>i}(\tau)$ denote a subsystem truncated BCH formula,
\begin{equation}
    \tilde{T}_{p,p_0}^{>i}(\tau) = \exp \left( -i \Phi^{>i}_{p,p_0}[\tau] \right), \quad \Phi^{>i}_{p,p_0}[\tau] = H^{>i}\tau + \sum_{q=2}^{p_0} \Phi_q^{>i} \tau^q,
\end{equation}
where $\Phi_q^{>i}$ is obtained by replacing $H_\gamma$ by $H_\gamma^{>i}$ in the BCH coefficient $\Phi_q$ [See Eq. (\ref{Eq_Pf:BCH})] for the order-$p$ Trotterization $T_p(\tau)$.
When we expand the operator $\tilde{T}_{p,p_0}^{>i}(\tau)^\dagger \tilde{T}_{p,p_0}^{>i-1}(\tau)$ by
\begin{equation}
    \tilde{T}_{p,p_0}^{>i}(\tau)^\dagger \tilde{T}_{p,p_0}^{>i-1}(\tau) = \sum_{q=0}^\infty \tilde{T}_q^i \tau^n,
\end{equation}
each coefficient is bounded by
\begin{equation}\label{EqA_Pf:tilde_Tn_bound}
    \| \tilde{T}_q^i \| \leq \frac{e^2}2 (8e^2 c_p p_0 kg)^q.
\end{equation}

\end{lemma}

\textbf{Proof.---} We define the difference $\Psi_{p,p_0}^i[\tau] \equiv \Phi^{>i}_{p,p_0}[\tau] - \Phi^{>i-1}_{p,p_0}[\tau]$.
Then, we use the relation by the Dyson series expansion, Eq. (\ref{Eq_Pf:Dyson}), for operators $A=-i\Phi_{p,p_0}^{>i}[\tau]$ and $B=i\Psi_{p,p_0}^i[\tau]$.
This leads to
\begin{eqnarray}
    && \tilde{T}_{p,p_0}^{>i}(\tau)^\dagger \tilde{T}_{p,p_0}^{>i-1}(\tau) \nonumber \\
    && \quad = e^{i\Phi_{p,p_0}^{>i}[\tau]}e^{-i\Phi_{p,p_0}^{>i}[\tau]+i\Psi_{p,p_0}^i[\tau]}  \nonumber \\
    && \quad =  \sum_{n=0}^\infty \int_0^1 \dd s_1 \int_0^{s_1} \dd s_2 \cdots \int_0^{s_{n-1}} \dd s_n \prod_{n'=1,\cdots,n}^\rightarrow \left( \sum_{l=0}^\infty \frac{(is_{n'})^l}{l!} \left( \ad_{\Phi_{p,p_0}^{>i}[\tau]} \right)^l i\Psi_{p,p_0}^i[\tau]\right). \nonumber \\
    &&\label{EqA_Pf:subsys_BCH_exp_Psi}
\end{eqnarray}
We define the series expansion of the term in the integral,
\begin{equation}\label{EqA_Pf:subsys_BCH_Phi}
    \sum_{l=0}^\infty \frac{(is)^l}{l!} \left( \ad_{\Phi_{p,p_0}^{>i}[\tau]}\right)^l i \Psi^i_{p,p_0}[\tau] = \sum_{q=0}^\infty \tilde{\Psi}_q(s) \tau^q,
\end{equation}
and then the coefficient $\tilde{T}_q^i$ is expressed by
\begin{equation}\label{EqA:tilde_T_n_product}
    \tilde{T}_q^i = \sum_{n=0}^\infty \int_0^1 \dd s_1 \int_0^{s_1} \dd s_2 \cdots \int_0^{s_{n-1}} \dd s_n \sum_{\substack{q_1,\cdots,q_n \geq 0:\\ q_1+\cdots+q_n=q}} \prod_{n'=1,\cdots,n}^\rightarrow \tilde{\Psi}_{q_{n'}}(s_{n'}).
\end{equation}
It suffices to evaluate the bound on $\| \tilde{\Psi}_q (s)\|$.

The difference $\Psi_{p,p_0}^i[\tau] = \Phi^{>i}_{p,p_0}[\tau] - \Phi^{>i-1}_{p,p_0}[\tau]$ is expanded by
\begin{equation}
    \Psi_{p,p_0}^i[\tau] = \sum_\gamma h_i^\gamma \tau + \sum_{q=2}^{p_0} \left( \Phi_q^{>i} - \Phi_q^{>i-1} \right) \tau^q \equiv \sum_{q=1}^{p_0} \Psi_q^i \tau^q.
\end{equation}
In a similar manner to Eq. (\ref{EqA:tilde_T_n_product}), the order-$q$ coefficient $\tilde{\Psi}_q(s)$ in Eq. (\ref{EqA_Pf:subsys_BCH_Phi}) is written by the products of $\ad_{\Phi_{q_{l'}}^{>i}}$ and $\Psi_{q'}^i$ satisfying $q'+q_1+\cdots+q_l=q$, which results in
\begin{equation}\label{EqA_Pf:tilde_Psi_bound}
    \| \tilde{\Psi}_q (s)\| \leq \sum_{l=0}^\infty \frac{1}{l!} \sum_{\substack{0 \leq q',q_1,\cdots,q_l \leq p_0: \\ q'+q_1+\cdots+q_l=q}} \norm{\left(\prod_{l'=1}^l \ad_{\Phi_{q_{l'}}^{>i}} \right)\Psi_{q'}^i}. 
\end{equation}
Let us denote the decomposition of $\Phi_q^{>i}$ into local terms by $\Phi_q^{>i} = \sum_X \phi_X^q$.
For $q'=1$, the norm of the nested commutator in the above formula is bounded by
\begin{eqnarray}
    && \norm{\left(\prod_{l'=1}^l \ad_{\Phi_{q_{l'}}^{>i}} \right)\Psi_{1}^i} \nonumber \\
    && \quad \leq \sum_{X: X \ni i} \sum_\gamma \sum_{X_1,\cdots,X_l} \norm{[\phi_{X_l}^{q_l},\cdots,[\phi_{X_1}^{q_1},h_X^\gamma]]} \nonumber \\
    && \quad \leq 2^l \sum_{X: X \ni i} \sum_\gamma \norm{h_X^\gamma} \sum_{X_1: X_1 \cap X \neq \phi} \sum_{X_2: X_2 \cap (Y_1 \cup X) \neq \phi} \cdots \sum_{X_l: X_l \cap (Y_{l-1}\cup X) \neq \phi} \prod_{l'=1}^l \norm{\phi_{X_{l'}}^{q_{l'}}}, \nonumber \\
    &&
\end{eqnarray}
where $Y_{l'}$ denotes the union $Y_{l'} = X_1 \cup \cdots \cup X_{l'}$ like Eq. (\ref{EqA_A0:proof1}).
Using the locality and extensiveness of $\Phi_q$ by Lemma \ref{LemmaA:locality_extensive_BCH} in a similar manner to Eq. (\ref{EqA_A0:proof2}), we obtain the upper bound,
\begin{eqnarray}
    \norm{\left(\prod_{l'=1}^l \ad_{\Phi_{q_{l'}}^{>i}} \right)\Psi_{1}^i} &\leq& 2^l g \prod_{l'=1}^l \left[ \left( \sum_{l''<l'} q_{l''}k + k\right) g(\Phi_{q_l'}) \right] \nonumber \\
    &\leq& 2^l g \prod_{l'=1}^l \left[ qk \frac{(q_{l'}-1)!}{q_{l'}} (2c_pkg)^{q_{l'}-1} c_pg \right] \nonumber \\
    &\leq& q^l g (2c_p p_0kg)^{q-1} \leq q^l (2c_p p_0kg)^q. \label{EqA_Pf:ad_Psi_l_1}
\end{eqnarray}
We used $\sum_{l''<l'} q_l'' + 1 \leq \sum_{l'=1}^l q_{l'} + q' = q$ in the second line and used $(q_{l'}-1)!/q_{l'} \leq p_0^{q_{l'}}$ in the third line, both of which are validated in the summation appearing in Eq. (\ref{EqA_Pf:tilde_Psi_bound}).

We next calculate the bounds of the terms with $q' \geq 2$ in Eq. (\ref{EqA_Pf:tilde_Psi_bound}).
The order-$q'$ term $\Psi_{q'}^i = \Phi_{q'}^{>i} - \Phi_{q'}^{>i-1}$ is expressed by
\begin{eqnarray}
    \Psi_{q'}^i &=& \sum_{\substack{q_1,\cdots,q_{c_p \Gamma} \geq 0: \\ q_1+\cdots+q_{c_p \Gamma} = q'}} (-i)^{q'-1} \left( \prod_{v=1}^{c_p\Gamma} \frac{(\alpha_v)^{q_v}}{q_v!}\right) \nonumber \\
    && \qquad \qquad\qquad \qquad\qquad \times \left[ \phi_{q'} (\underbrace{H_1^{>i},\cdots,H_{c_p\Gamma}^{>i}}_{q'}) - \phi_{q'} (\underbrace{H_1^{>i-1},\cdots,H_{c_p\Gamma}^{>i-1}}_{q'}) \right] \nonumber \\
    &=& \sum_{\substack{q_1,\cdots,q_{c_p \Gamma} \geq 0: \\ q_1+\cdots+q_{c_p \Gamma} = q'}} (-i)^{q'-1} \left( \prod_{v=1}^{c_p\Gamma} \frac{(\alpha_v)^{q_v}}{q_v!}\right) \nonumber \\
    && \quad \times \left[ \phi_{q'}(h_i^1,\underbrace{H_1^{>i}, \cdots, H_1^{>i}}_{q_1-1},  \cdots ,\underbrace{H_{c_p\Gamma}^{>i}, \cdots, H_{c_p\Gamma}^{>i}}_{q_{c_p\Gamma}}) \right. \nonumber \\
    && \qquad \qquad + \phi_{q'}(H_1^{>i-1},h_i^1,\underbrace{H_1^{>i}, \cdots, H_1^{>i}}_{q_1-2},  \cdots ,\underbrace{H_{c_p\Gamma}^{>i}, \cdots, H_{c_p\Gamma}^{>i}}_{q_{c_p\Gamma}}) + \cdots   \nonumber \\
    && \qquad \qquad \qquad \left. \cdots + \phi_{q'}(\underbrace{H_1^{>i-1}, \cdots, H_1^{>i-1}}_{q_1},  \cdots ,\underbrace{H_{c_p\Gamma}^{>i-1}, \cdots, H_{c_p\Gamma}^{>i-1}}_{q_{c_p\Gamma}-1}, h_i^{c_p\Gamma}) \right], \nonumber \\
    && \label{EqA_Pf:Psi_q_local}
\end{eqnarray}
The norm appearing in Eq. (\ref{EqA_Pf:tilde_Psi_bound}) can be evaluated in a similar manner to the proofs in Lemmas \ref{LemmaA:nested_bound} and \ref{LemmaA:locality_extensive_BCH}.
For instance, when the nested commutator $\prod_{l'} \ad_{\Phi_{q_{l'}}^{>i}}$ is applied to the first term in Eq. (\ref{EqA_Pf:Psi_q_local}), we have
\begin{eqnarray}
    && \norm{\sum_{\substack{q_1,\cdots,q_{c_p \Gamma}: \\ q_1+\cdots+q_{c_p \Gamma} = q'}} (-i)^{q'-1} \left( \prod_{v=1}^{c_p\Gamma} \frac{(\alpha_v)^{q_v}}{q_v!}\right) \left( \prod_{l'=1}^l \ad_{\Phi_{q_{l'}}^{>i}} \right) \phi_{q'}(h_i^1,H_1^{>i}, \cdots, H_{c_p\Gamma}^{>i})} \nonumber \\
    && \quad \leq \frac{(c_p)^q}{q^2} \max_{\sigma \in S_q}  \sum_{\substack{X_1,\cdots,X_q: \\ X_1 \ni i}} \sum_{X_{q+1},\cdots,X_{q+l}}\norm{[\phi_{X_{q+l}}^{q_l}, \cdots,[\phi_{X_{q+1}}^{q_1},[h_{X_{\sigma(1)}},\cdots,[h_{X_{\sigma(q-1)}}, h_{X_{\sigma(q)}}]]]} \nonumber \\
    && \quad \leq 2^l \frac{(c_p)^{q'}}{{q'}^2} \max_{\sigma \in S_{q'}}  \sum_{\substack{X_1,\cdots,X_{q'}: \\ X_1 \ni i}} \norm{[h_{X_{\sigma(1)}},\cdots,[h_{X_{\sigma(q'-1)}}, h_{X_{\sigma(q')}}]]} \nonumber \\
    && \qquad \qquad\qquad \qquad\qquad \times \sum_{\substack{X_{q'+1}:\\ X_{q'+1} \cap Y_{q'} \neq \phi}} \norm{\phi_{X_{q'+1}}^{q_1}} \cdots \sum_{\substack{X_{q'+l}:\\ X_{q'+l} \cap Y_{q'+l-1} \neq \phi}} \norm{\phi_{X_{q'+l}}^{q_l}}.\label{Eq_A0:Psi_q_local_first}
\end{eqnarray}
We note that the superscript ``$>i$'' is not significant here since the locality and extensiveness are respectively maintained.

\begin{eqnarray}
    [\text{Eq. (\ref{Eq_A0:Psi_q_local_first})}] &\leq& 2^l \frac{(c_p)^{q'}}{q'^2} \sum_{X_1: X_1 \ni i} (q'-1)! (2kg)^{q'-1} \norm{h_{X_1}} \prod_{l'=1}^l \left[ \left( \sum_{l''<l'} q_{l''}k+ q'k\right) g(\Phi_{q_{l'}})\right] \nonumber \\
    &\leq& \frac{q^l}{q'^2} (2c_p p_0kg)^{q},
\end{eqnarray}
where we used the condition, $q'+q_1+\cdots+q_l =q$ in the last inequality.
The nested commutators with the other $q'-1$ terms in Eq. (\ref{EqA_Pf:Psi_q_local}) share the same bound, which results in
\begin{equation}\label{EqA_Pf:ad_Psi_q_bound}
    \norm{\left(\prod_{l'=1}^l \ad_{\Phi_{q_{l'}}^{>i}} \right)\Psi_{q'}^i} \leq q' \times \frac{q^l}{q'^2} (2c_p p_0 kg)^{q} \leq q^l (2 c_p p_0kg)^q.
\end{equation}
also for $q' \geq 2$.

Summarizing the results of Eqs. (\ref{EqA_Pf:ad_Psi_l_1}) and (\ref{EqA_Pf:ad_Psi_q_bound}), the coefficient appearing the series expansion Eq. (\ref{EqA_Pf:tilde_Psi_bound}) is bounded by
\begin{eqnarray}
    \|\tilde{\Psi}_q(s)\| &\leq& \sum_{l=0}^\infty \frac{1}{l!}  \sum_{\substack{q',q_1,\cdots,q_l \geq 0: \\ q'+q_1+\cdots+q_l=q}} q^l (2c_p p_0 kg)^q \nonumber \\
    &\leq& (4c_p p_0 kg)^q \sum_{l=0}^\infty \frac{(2q)^l}{l!} = (4e^2 c_p p_0 kg)^q.
\end{eqnarray}
We used the relation $\sum_{\substack{q',q_1,\cdots,q_l \geq 0: \\ q'+q_1+\cdots+q_l=q}}1 = (q+l)!/q!l!\leq 2^{q+l}$ in the second inequality.
Finally, substituting this into Eq. (\ref{EqA:tilde_T_n_product}), we obtain
\begin{eqnarray}
    \| \tilde{T}_q^i \| &\leq& \sum_{n=0}^\infty \int_0^1 \dd s_1 \cdots \int_0^{s_{n-1}} \dd s_n \sum_{\substack{q_1,\cdots,q_n \geq 0: \\ q_1+\cdots+q_n=q}} \prod_{n'=1}^n \| \tilde{\Psi}_{q_{n'}}(s_{n'})\| \nonumber \\
    &\leq& \sum_{n=0}^\infty \frac1{n!} \sum_{\substack{q_1,\cdots,q_n \geq 0: \\ q_1+\cdots+q_n=q}} (4e^2 c_p p_0 kg)^{q_1+\cdots+q_n} \nonumber \\
    &\leq& e^2 2^{q-1} (4e^2 c_p p_0 kg)^q,
\end{eqnarray}
which completes the proof of Eq. (\ref{EqA_Pf:tilde_Tn_bound}). $\quad \square$

\subsection{Improvements from the proof for the Floquet-Magnus expansion}\label{SubsecA:improvement}
Trotterization can be seen as the time-evolution operator under the time-periodic Hamiltonian $H^{>0}(\tau')$ by Eq. (\ref{EqA:Trotter_periodic}), which is piecewise constant and has the period $c_p\Gamma \tau$.
Compared to the original proof for generic time-periodic Hamiltonians \cite{Kuwahara2016-yn}, our results by Lemma \ref{LemmaA:Series_subsys_Trot}, Lemma \ref{LemmaA:Series_subsys_BCH}, and Theorem \ref{Thm:error_truncated_BCH} are better by the factor $\Gamma$ [the number of partitions in Trotterization by Eq. (\ref{Eq_Pre:H_partition})].
To be precise, the original proof predicts
\begin{equation}
    \norm{T_q^i} \in \order{(\Gamma kg)^q}, \quad \| \tilde{T}_q^i \| \in \order{(\Gamma p_0 kg)^q}
\end{equation}
as counterparts of Lemmas \ref{LemmaA:Series_subsys_Trot} and \ref{LemmaA:Series_subsys_BCH}, coming from the period $c_p\Gamma \tau$.
Accompanied by this additional factor, it is demanded that the time $\tau$ is small enough to satisfy $\tau \in \order{(\Gamma p_0kg)^{-1}}$ in order to approximate Trottetization by the truncated BCH formula as Eq. (\ref{Eq_Pf:error_truncated_BCH}).
On the other hand, we exploit the structure of Trotterization, Eq. (\ref{Eq_Pre:Trotterization}), especially about the nested commutator Eq. (\ref{Eq_Pre:commutator_bound}).
This leads to the better bound in Lemmas \ref{LemmaA:Series_subsys_Trot} and \ref{LemmaA:Series_subsys_BCH}, and the much looser requirement $\tau \in \order{(p_0 kg)^{-1}}$ by Eq. (\ref{Eq_Pf:time_condition}), which is independent of $\Gamma$.

We note that this improvement is essential in Hamiltonian simulation, especially for Hamiltonians with long-range interactions.
For generic Hamiltonians with finite-range interactions, the number of terms $\Gamma$ can be $\order{1}$ and hence it seems not to be important.
However, when limiting to finite-range interactions, the Haah-Hastings-Kothari-Low (HHKL) algorithm can achieve nearly optimal gate counts $\order{Ngt \times \mr{polylog}(Ngt/\varepsilon)}$, which is better than MPF \cite{Haah2021-time-dep} (See also Table \ref{Tab:Gate_counts}).
There is stronger motivation in considering long-range interactions for MPF, where the HHKL algorithm becomes inefficient and the number of terms $\Gamma$ generally increases in the number of interactions, $\order{N^k}$.
In that case, the original result for generic cases demands the Trotter step $\tau = t/r \in \order{(\Gamma p_0kg)^{-1}}$.
This is not compatible with the good size-dependency by Eq. (\ref{Eq_Pf:MPF_complexity}).
Namely, it is essential to delete the $\Gamma$-dependence from the requirement as Theorem \ref{Thm:error_truncated_BCH}.
Our improvement can ensure that MPF has a cost with good size-dependency, Eq. (\ref{Eq_Pf:MPF_complexity}), broadly for finite-ranged and long-ranged interacting Hamiltonians.

\end{document}